\documentclass[12pt]{article}
\usepackage{amsfonts}

\newcommand{\newsection}{    % Numeration of eqs. is automatic
\setcounter{equation}{0}\section}
\def\appendix#1{\addtocounter{section}{1}\setcounter{equation}{0}
\renewcommand{\thesection}{\Alph{section}}
\section*{Appendix \thesection\protect\indent \parbox[t]{11.15cm}{#1}}
\addcontentsline{toc}{section}{Appendix \thesection\ \ \ #1}}

\font\mybb=msbm10 at 11pt

\def\bb#1{\hbox{\mybb#1}}

\def\bR {\bb{R}}

\begin{document}
\begin{titlepage}
\begin{center}
%\today
\vspace*{-1.0cm}
%\hfill hep-th/yymmnnn \\
%\hfill UB-ECM-PF-06-43 \\

\vspace{2.0cm} {\Large \bf  IIB black hole horizons with five-form flux and KT geometry} \\[.2cm]

\vspace{1.5cm}
 {\large  U. Gran$^1$, J. Gutowski$^2$ and  G. Papadopoulos$^2$}

\vspace{0.5cm}

${}^1$ Fundamental Physics\\
Chalmers University of Technology\\
SE-412 96 G\"oteborg, Sweden\\
%\vspace{0.5cm}

\vspace{0.5cm}
${}^2$ Department of Mathematics\\
King's College London\\
Strand\\
London WC2R 2LS, UK\\

\vspace{0.5cm}

\end{center}

\vskip 1.5 cm
\begin{abstract}
We investigate the near horizon  geometry of IIB supergravity black holes with non-vanishing 5-form flux preserving at least
two supersymmetries. We demonstrate that there are three classes of solutions distinguished by the choice of Killing spinors.
We find that the  spatial horizon sections of the class of solutions with an $SU(4)$ invariant pure Killing spinor are hermitian manifolds
 and admit a hidden K\"ahler with torsion (KT) geometry  compatible with the $SU(4)$
structure. Moreover the Bianchi identity of the 5-form, which also implies the field equations,
 can be expressed in terms of the torsion $H$ as $d(\omega\wedge H)=\partial\bar\partial \omega^2=0$,
where $\omega$ is a Hermitian form. We give several examples of near horizon geometries which include group manifolds, group fibrations over KT manifolds
and uplifted geometries of lower dimensional black holes. Furthermore, we show that the class of solutions associated with a $Spin(7)$ invariant spinor
is locally a product $\bR^{1,1}\times {\cal S}$, where ${\cal S}$ is a holonomy $Spin(7)$ manifold.

\end{abstract}

\end{titlepage}

\section{Introduction}

There is  evidence to suggest  that in higher dimensions there are black holes with
exotic horizon topologies. As a result, the classical black hole uniqueness theorems \cite{israel}-\cite{robinson}
do not extend to more than four dimensions. Five dimensions are also special. Although there is no uniqueness theorem
 for a large class of theories
the horizon topologies that can occur
are $S^3$, $S^1\times S^2$ and $T^3$ \cite{reall}. The first two are the horizon topologies of the BMPV black hole \cite{bmpv} and black ring \cite{ring1},
respectively. To our knowledge, no black hole solution has been found with horizon topology $T^3$.

To probe the horizon topologies in more than five dimensions, one can either assume that the solutions are static, see e.g.~\cite{gibbons1,rogatkov1, rogatkov2}, or are black hole solutions of the type considered in \cite{obers1a, obers1b},
or that they preserve a fraction of spacetime supersymmetry. The latter assumption is natural in the context of string,
Kaluza-Klein or supergravity theories.  The analysis is further simplified provided that one
considers extreme black holes and focuses on a suitable  geometry near the horizon, the near horizon geometry\footnote{However, it is not apparent that all near horizon geometries found in such an investigation can be extended to full black hole solutions. For an extensive discussion on this point, see eg \cite{hhv1} and references within.}.
In this context,
it is natural to ask whether the topology and geometry of the near horizon geometries
of supersymmetric black holes of higher-dimensional supergravity theories can be classified.
Some progress has been made to solve this problem. For example, there is a good understanding of
the near horizon topologies and geometries of heterotic supergravity \cite{hhv1, hhv2}. This has been assisted
by the  solution of the Killing spinor equations (KSEs) of heterotic supergravity in all cases \cite{hetv1, hetv2, hetv3}.
 In particular, all the conditions on the geometry of  heterotic horizons are known, as well as the corresponding
  fractions of supersymmetry  preserved.
 The half supersymmetric horizons
have been classified, and the $1/4$ supersymmetric ones lead to pairing of a cohomological and of a non-linear differential system
 on K\"ahler surfaces.  Although there is no classification of the $1/4$ supersymmetric horizons,
many explicit solutions of both systems are known, for example on del Pezzo surfaces, and the associated
horizons have exotic topologies.

In this paper we extend the results of the heterotic analysis to type IIB supergravity \cite{west, schwarz, howe}.
In contrast to the heterotic case,  somewhat
less is known about solutions of IIB supergravity.
In particular, the KSEs have been solved for $N=1$ backgrounds in \cite{sgiibv1, sgiibv2}. It has also been shown that
if a background preserves more than 28 supersymmetries it is maximally supersymmetric \cite{iibm28v1, iibm28v2}. Moreover, the backgrounds that
preserve 28 and 32 supersymmetries have been classified in \cite{iib28} and \cite{iib32}, respectively. Very little is known
about the properties  of solutions in the intermediate cases, however see the conjectures in \cite{duff, jose}.
Some simplification occurs for those backgrounds
that have only 5-form flux \cite{iibfive}. Because of this, we shall first examine  the supersymmetric IIB near horizon
geometries with non-vanishing 5-form flux. The general case which includes IIB near horizon geometries with other fluxes
will be reported elsewhere. The advantage of focusing on near horizon geometries with only 5-form flux is that
the analysis is rather economical and leads to insightful connections with KT geometry. This in turn
allows for the construction of many examples of near horizon geometries, some of which have exotic topologies.

The focus of our analysis is on  IIB near horizon geometries with non-vanishing 5-form flux that
preserve at least 2 supersymmetries\footnote{ This is the minimal amount of supersymmetry that is preserved by a solution when only the 5-form flux is non-vanishing.}.
An application of the spinorial geometry technique  \cite{spingem} for solving KSEs  to IIB supergravity  reveals
that there are three classes of near horizon geometries depending on the choice of Killing spinors. The Killing spinor of the
 first class of solutions is constructed from a
 $Spin(7)$ invariant spinor on the spatial horizon section ${\cal S}$. In this case we shall show that the near horizon geometry is $\mathbb{R}^{1,1}\times {\cal S}$.  In turn
${\cal S}$ is  a product of closed Riemannian manifolds
with special holonomy as  given in the Berger classification, and the 5-form vanishes.
The Killing spinors of the other two classes are constructed from $SU(4)$ invariant spinors on ${\cal S}$ .
These two classes are distinguished by whether the $SU(4)$ invariant
 spinors are generic or pure. We shall focus our analysis on the pure case. The geometry of the horizons in the  generic $SU(4)$ case is different
and its exploration requires the development of new techniques
which will be reported elsewhere.

The  Killing spinor vector bi-linear of the pure $SU(4)$  invariant case, which we identify with the
black hole stationary Killing vector field, is null. Consequently, the metric of the near horizon geometry can be written
as
\begin{eqnarray}
ds^2=2 du (dr + rh )+ ds^2_{(8)}({\cal S})~,
\end{eqnarray}
where $ds^2_{(8)}({\cal S})$ is the metric of the horizon section. Moreover, the KSEs require that
 ${\cal S}$ is a Hermitian manifold with an $SU(4)$ structure
such that
\begin{eqnarray}
h=\theta_\omega=\theta_{\mathrm{Re}\,\chi}~,
\end{eqnarray}
where $\theta_\omega$ and $\theta_{\mathrm{Re}\,\chi}$ are  the Lee forms of the Hermitian form $\omega$ and the real component of the
(4,0)-form $\chi$, respectively.

The equality of the two Lee forms is significant. This is because it is precisely
 the condition for the $SU(4)$ structure on ${\cal S}$ to admit a compatible K\"ahler  with torsion (KT)  geometry \cite{hkt}.
 This condition implies that the manifold is equipped
 with a metric connection, $\hat\nabla$,
  with skew-symmetric torsion $H$, such that
 \begin{eqnarray}
 \hat\nabla \omega=\hat\nabla \chi=0~.
 \label{lee}
 \end{eqnarray}
 Therefore all the horizon sections
 with non-vanishing 5-form flux admit a {\it hidden} 3-form torsion. This cannot be immediately identified with either
 the NS-NS 3-form or R-R field strengths of IIB supergravity as they have been set to zero. Another advantage of
 introducing $H$ is that now the Bianchi identity for the 5-form, which also implies all the remaining equations
 of IIB supergravity including field equations, can be written as
 \begin{eqnarray}
 d(\omega\wedge H)=i\partial\bar\partial\omega^2=0~.
 \label{oskt}
 \end{eqnarray}
 As we shall demonstrate, expressing the conditions implied by the KSE and field equations as in (\ref{lee}) and (\ref{oskt})
  is instrumental for the construction of many examples of near horizon geometries.
Our examples include horizons with sections which are group manifolds, and toric and $SU(2)$ fibrations over
 lower dimensional KT manifolds.
A particular large class of examples includes $T^2$ fibrations over 6-dimensional K\"ahler-Einstein
manifolds. We also demonstrate that the uplifting of the near horizon geometries of 5-dimensional black holes \cite{adsfive, chong, reallk}
to IIB solves all the conditions and so provides more examples.

 IIB spatial horizon sections are 8-dimensional   but the conditions we have found on ${\cal S}$ can be easily
adapted to 2n dimensions.  Strong KT manifolds (SKT) \cite{hkt} are KT manifolds which in addition satisfy the second order
equation $dH=2i\partial\bar\partial\omega=0$.
A comparison of  (\ref{oskt}) with the strong condition for SKT manifolds
   leads to a generalization of both conditions. In particular, k-strong K\"ahler manifolds with torsion  (k-SKT)   are
   KT manifolds which in addition satisfy
   $\partial\bar\partial\omega^k=d(\omega^{k-1}\wedge H)=0$.
For 2n-dimensional manifolds, the (n-1)-SKT and (n-2)-SKT structures coincide with the Gauduchon \cite{gauduchon} and the Jost and Yau
 astheno-K\"ahler \cite{jost} conditions, respectively.
The above conditions can also be extended to 2n-dimensional manifolds with an $SU(n)$ structure compatible with a connection
with skew-symmetric torsion, or equivalently almost Calabi-Yau with torsion (ACYT)  and, if the almost complex structure is integrable,
Calabi-Yau with torsion (CYT) manifolds.
In this terminology, the horizon spatial section ${\cal S}$ is a 2-SCYT manifold.
The expression of k-SKT structure  in terms
of $H$ allows one to further extend it  on other manifolds with almost KT (AKT),
 $Sp(n)$, $Sp(n)\cdot Sp(1)$,  $G_2$ or $Spin(7)$ structures.

 A further generalization of k-SKT geometries is possible following the introduction of the k-Gauduchon condition
 $\partial\bar\partial\omega^k\wedge \omega^{n-k-1}=0$ for Hermitian manifolds in \cite{fino3}. One
 can also define the $(k;\ell)$-SKT condition as $\partial\bar\partial \omega^k\wedge \omega^\ell=0$ which includes both the
 k-SKT and k-Gauduchon structures. Rewriting this as $d(\omega^{k-1}\wedge H)\wedge \omega^\ell=0$
 it generalizes to other manifolds with $SU(n)$, $Sp(n)$ and $Sp(n)\cdot Sp(1)$ structures.

This paper is organized as follows. In section two, we describe the field and KSEs for near horizon geometries of IIB supergravity. In section 3,
we solve the KSEs for horizons which preserve at least two supersymmetries. The cases with $Spin(7)$-invariant and pure $SU(4)$-invariant Killing spinors
are emphasized. In section 4, we demonstrate that the spatial horizon sections of solutions  with a pure $SU(4)$-invariant Killing spinor admit
a hidden KT geometry compatible with an $SU(4)$ structure. In section 5, we give several examples of IIB supersymmetric horizons which are group
fibrations over KT manifolds. In section 6, we present some more examples which arise  by  uplifting lower-dimensional black hole horizons
to IIB supergravity, and in section 7, we give our conclusions. In appendix A, we explain our conventions.  In appendix B we give the definitions
of new geometries associated with other structure groups which arise as a generalization of the conditions we have found on the IIB spatial  horizon sections. In appendix
C, we give the 5-form field strength of the uplifted lower-dimensional black hole horizon geometries.

\newsection{Fields near the horizon and supersymmetry}

\subsection{Near horizon limit and field equations}

It is well-known that under some analyticity assumptions \cite{wald}, one can adapt  Gaussian Null co-ordinates near the horizon of
 an extremal black hole to write the metric as
 \begin{eqnarray}
ds^2 =2 {\bf{e}}^+ {\bf{e}}^- + \delta_{ij} {\bf{e}}^i {\bf{e}}^j = 2 du (dr + rh -{1 \over 2} r^2 \Delta du)+ \gamma_{IJ} dy^I dy^J~,
\end{eqnarray}
where we have introduced the basis
\begin{eqnarray}
\label{basis1}
{\bf{e}}^+ = du, \qquad {\bf{e}}^- = dr + rh -{1 \over 2} r^2 \Delta du, \qquad {\bf{e}}^i = e^i_I dy^I~.
\end{eqnarray}
The horizon is the Killing horizon of the time-like Killing vector field  $V={\partial \over \partial u}$ which is identified
to be in the same class as the stationary Killing vector field of the black hole, see e.g.~\cite{wald}, \cite{hhv1}. The spatial horizon section ${\cal S}$
is the co-dimension 2 submanifold defined by $r=u=0$ and it is assumed to be closed, i.e.~compact without boundary.

The components of the metric depend on all coordinates apart from $u$. The near horizon geometry is defined
by first making the coordinate transformation
\begin{eqnarray}
r \rightarrow \ell r, \qquad u \rightarrow \ell^{-1} u~,
\end{eqnarray}
and then taking the limit $\ell \rightarrow 0$. The resulting spacetime metric does not change its form, however
in the near-horizon limit $\Delta$, $h$ and $\gamma$ no longer depend on $r$.
The components of the spin connection are listed in  Appendix A.

The self-dual\footnote{In our conventions $F_{M_1 \dots M_5} =
{1 \over 5!} \epsilon^{N_1 \dots N_5}{}_{M_1 \dots M_5}
F_{N_1 \dots N_5}$, where $\epsilon_{0123456789}=1$.}
5-form field strength $F$ of IIB supergravity also simplifies in the near horizon limit. Assuming  that all
components of $F$ are regular functions of $r$, independent of $u$, such that
$F$ is well-defined on taking the near-horizon limit, in addition to the duality condition
and the Bianchi identity $dF=0$, one finds that
\begin{eqnarray}
\label{fivef}
F= r du \wedge dY + du \wedge dr \wedge Y - \star_8 Y = r {\bf{e}}^+ \wedge (dY-h \wedge Y)
+ {\bf{e}}^+ \wedge {\bf{e}}^- \wedge Y - \star_8 Y~,
\end{eqnarray}
where $Y$ is a $r,u$-independent 3-form on ${\cal{S}}$. Writing the 10-dimensional spacetime volume form
 in terms of
that on ${\cal{S}}$  as
\begin{eqnarray}
d{\rm vol}_{(10)} = {\bf{e}}^+ \wedge {\bf{e}}^- \wedge d{\rm vol}_{(8)}~,
\end{eqnarray}
one finds that $Y$ satisfies
\begin{eqnarray}
\label{sd1}
d \star_8 Y=0, \qquad dY-h \wedge Y = - \star_8 (dY-h \wedge Y )~,
\end{eqnarray}
and $\star_8$ is the Hodge dual on ${\cal{S}}$, with the convention that
\begin{eqnarray}
(\star_8 Y)_{n_1 n_2 n_3 n_4 n_5} = {1 \over 3!} \epsilon^{m_1 m_2 m_3}{}_{n_1 n_2 n_3 n_4 n_5}Y_{m_1 m_2 m_3}~.
\end{eqnarray}

The field equation for the 5-form field strength coincides with the Bianchi identity which we have already given
in (\ref{sd1}). The remaining field equation is the Einstein equation of the theory,
\begin{eqnarray}
R_{AB} = {1 \over 6} F_{A L_1 L_2 L_3 L_4} F_B{}^{L_1 L_2 L_3 L_4}~.
\end{eqnarray}
For the near horizon geometry, this can be decomposed along the light-cone directions
and those of the horizon section ${\cal S}$.   In particular, from the $+-$ component, one obtains:
\begin{eqnarray}
\label{ein1}
{1 \over 2} {\tilde{\nabla}}^i h_i - \Delta - {1 \over 2} h^2 = -{2 \over 3} Y_{\ell_1 \ell_2 \ell_3} Y^{\ell_1 \ell_2 \ell_3}~,
\end{eqnarray}
where ${\tilde{\nabla}}$ denotes the Levi-Civita connection of ${\cal{S}}$. From the $ij$ component one finds
\begin{eqnarray}
\label{ein2}
{\tilde{R}}_{ij} + {\tilde{\nabla}}_{(i} h_{j)} -{1 \over 2} h_i h_j = -4 Y_{i \ell_1 \ell_2} Y_j{}^{\ell_1 \ell_2}+{2 \over 3} \delta_{ij}
Y_{n_1 n_2 n_3}Y^{n_1 n_2 n_3}~,
\end{eqnarray}
where ${\tilde{R}}$ denotes the Ricci tensor of ${\cal{S}}$. From the $++$ component, one obtains
\begin{eqnarray}
\label{ein3}
{1 \over 2} {\tilde{\nabla}}^2 \Delta -{3 \over 2} h^i {\tilde{\nabla}}_i \Delta -{1 \over 2} \Delta {\tilde{\nabla}}^i h_i + \Delta h^2
+{1 \over 4} dh_{ij} dh^{ij} \nonumber \\ = {1 \over 6} (dY-h \wedge Y)_{n_1 n_2 n_3 n_4} (dY-h \wedge Y)^{n_1 n_2 n_3 n_4}~,
\end{eqnarray}
and from the $+i$ component, one gets
\begin{eqnarray}
\label{ein4}
{1 \over 2} {\tilde{\nabla}}^j dh_{ij} - h^j dh_{ij} - {\tilde{\nabla}}_i \Delta + \Delta h_i = -{4 \over 3} (dY-h \wedge Y)_{i n_1 n_2 n_3} Y^{n_1 n_2 n_3}~.
\end{eqnarray}

\subsection{Killing spinor equations }

We set the axion and  dilaton to be constant, and the 3-forms to vanish.  Thus the only active
bosonic fields are  the metric and  real self-dual 5-form $F$. In such case, the only non-trivial KSE
is
\begin{eqnarray}
\label{kse}
\nabla_M\epsilon
+{i \over 48} F_{M N_1 N_2 N_3 N_4}\Gamma^{N_1 N_2 N_3 N_4}  \epsilon =0~,
\end{eqnarray}
where $\nabla$ is the spin connection associated with the frame (\ref{basis1}) and $\epsilon$ is a spinor in the positive chirality
complex Weyl representation of $Spin(9,1)$.

To proceed further, we use the projections
\begin{eqnarray}
\epsilon=\epsilon_++\epsilon_-~,~~~\Gamma_\pm \epsilon_\pm=0~,
\end{eqnarray}
to decompose the KSE along the light-cone directions and the rest. The KSE along the light-cone directions can be integrated.
In particular on integrating up the $-$ component of the KSE, one finds
\begin{eqnarray}
\label{st1}
\epsilon_+=\phi_+, \qquad \epsilon_-=\phi_- + r \Gamma_- \bigg({1 \over 4} h_i \Gamma^i
+{i \over 12} Y_{n_1 n_2 n_3} \Gamma^{n_1 n_2 n_3} \bigg) \phi_+~,
\end{eqnarray}
where $\phi_\pm$ do not depend on $r$. A similar analysis of the $+$ component of the KSE gives that
\begin{eqnarray}
\label{st2}
\phi_+ =\eta_+ + u \Gamma_+ \bigg({1 \over 4} h_i \Gamma^i
-{i \over 12} Y_{n_1 n_2 n_3} \Gamma^{n_1 n_2 n_3} \bigg) \eta_- , \qquad \phi_- = \eta_-~,
\end{eqnarray}
where $\eta_\pm$ do not depend on $u$ or $r$. Furthermore, $\eta_+, \eta_-$ must  satisfy the following algebraic
conditions

\begin{eqnarray}
\label{alg1}
\bigg(
-{1 \over 8} h^2 -{1 \over 2} \Delta +{1 \over 12} Y_{\ell_1 \ell_2 \ell_3} Y^{\ell_1 \ell_2 \ell_3}
-{1 \over 8} dh_{ij} \Gamma^{ij}
\nonumber \\
+\big({i \over 48} dY_{\ell_1 \ell_2 \ell_3 \ell_4}
-{1 \over 8}Y_{m \ell_1 \ell_2} Y^m{}_{\ell_3 \ell_4}\big)
\Gamma^{\ell_1 \ell_2 \ell_3 \ell_4} \bigg) \eta_-=0~,
\end{eqnarray}

\begin{eqnarray}
\label{alg2}
\bigg(
{1 \over 8} h^2 +{1 \over 2} \Delta -{1 \over 12} Y_{\ell_1 \ell_2 \ell_3} Y^{\ell_1 \ell_2 \ell_3}
-{1 \over 8} dh_{ij} \Gamma^{ij}
\nonumber \\
+\big({i \over 48} dY_{\ell_1 \ell_2 \ell_3 \ell_4}
+{1 \over 8}Y_{m \ell_1 \ell_2} Y^m{}_{\ell_3 \ell_4}\big)
\Gamma^{\ell_1 \ell_2 \ell_3 \ell_4} \bigg) \eta_+=0~,
\end{eqnarray}

\begin{eqnarray}
\label{alg3}
\bigg(
{1 \over 8} h^2 +{1 \over 2} \Delta -{1 \over 12} Y_{\ell_1 \ell_2 \ell_3} Y^{\ell_1 \ell_2 \ell_3}
-{1 \over 8} dh_{ij} \Gamma^{ij}
\nonumber \\
+\big({i \over 48} dY_{\ell_1 \ell_2 \ell_3 \ell_4}
+{1 \over 8}Y_{m \ell_1 \ell_2} Y^m{}_{\ell_3 \ell_4}\big)
\Gamma^{\ell_1 \ell_2 \ell_3 \ell_4} \bigg)
\nonumber \\ \times
\bigg({1 \over 4} h_j \Gamma^j -{i \over 12} Y_{n_1 n_2 n_3} \Gamma^{n_1 n_2 n_3} \bigg) \eta_-
=0~,
\end{eqnarray}

\begin{eqnarray}
\label{alg4}
\bigg( \big(-{1 \over 8} dh_{q_1 q_2} \Gamma^{q_1 q_2} +{i \over 48} (dY-h \wedge Y)_{\ell_1 \ell_2 \ell_3 \ell_4}
\Gamma^{\ell_1 \ell_2 \ell_3 \ell_4} \big) ({1 \over 4} h_j \Gamma^j
+{i \over 12} Y_{n_1 n_2 n_3} \Gamma^{n_1 n_2 n_3} ) \nonumber \\ +  {1 \over 4} (\Delta h_i - \partial_i \Delta) \Gamma^i \bigg) \eta_+=0~,
\end{eqnarray}
and
\begin{eqnarray}
\label{alg5}
\bigg( \big(-{1 \over 8} dh_{q_1 q_2} \Gamma^{q_1 q_2} +{i \over 48} (dY-h \wedge Y)_{\ell_1 \ell_2 \ell_3 \ell_4}
\Gamma^{\ell_1 \ell_2 \ell_3 \ell_4} \big) ({1 \over 4} h_j \Gamma^j
+{i \over 12} Y_{n_1 n_2 n_3} \Gamma^{n_1 n_2 n_3} ) \nonumber \\ +  {1 \over 4} (\Delta h_i - \partial_i \Delta) \Gamma^i \bigg)
\bigg({1 \over 4} h_m \Gamma^m -{i \over 12} Y_{m_1 m_2 m_3} \Gamma^{m_1 m_2 m_3} \bigg) \eta_-=0~.
\end{eqnarray}

It has been shown in \cite{sgiibv1, sgiibv2} that all supersymmetric IIB backgrounds admit a Killing vector
field constructed as a bilinear of the Killing spinor. The solution of the above algebraic conditions as well
as  that of the remaining component of the KSE along ${\cal S}$ proceeds by identifying the Killing vector bilinear
with the Killing vector field of the near horizon geometry $V=\partial_u$. This is justified if one assumes that
the black hole spacetime is supersymmetric. However, this is not necessary. As it has been emphasized in \cite{n1d4},
the analysis can be carried out  under the assumption that only the near horizon geometry is supersymmetric and not
necessarily the black hole spacetime. However, such a weaker assumption leads to a more involved analysis in
IIB supergravity which is not within the scope of this paper.

\newsection{Solutions with at least two supersymmetries}

To proceed, we consider first the solutions with minimal supersymmetry, and we require that
the 1-form Killing spinor bilinear
\begin{eqnarray}
Z_M = \langle B (C \epsilon^*)^*, \Gamma_M \epsilon \rangle
= \langle \Gamma_0 \epsilon, \Gamma_M \epsilon \rangle~,
\end{eqnarray}
should be proportional to $V$, where
\begin{eqnarray}
V = -{1 \over 2} r^2 \Delta {\bf{e}}^+ + {\bf{e}}^- \ .
\end{eqnarray}

First, evaluate $Z$ at $r=u=0$, for which $\epsilon=\eta_++ \eta_-$.
Requiring that $Z_+=0$ at $r=u=0$ implies that
\begin{eqnarray}
\eta_-=0 \ .
\end{eqnarray}
Then, using $r,u$ independent $Spin(8)$ gauge transformations of the type considered in \cite{sgiibv1, sgiibv2},
one can, without loss of generality, take
\begin{eqnarray}
\eta_+=p+q e_{1234}~,
\label{pqe}
\end{eqnarray}
where $p,q$ are complex functions of ${\cal S}$. Furthermore, on computing the component $Z_-$, one finds that
$|p|^2+|q|^2$ must be a (non-zero) constant.

Next, evaluate $Z_i$ at $r \neq 0$. As this component must vanish, one finds
\begin{eqnarray}
\label{hexp}
h_i =- {|p|^2-|q|^2 \over |p|^2+|q|^2} Y_{i \ell_1 \ell_2} \omega^{\ell_1 \ell_2}~,
\end{eqnarray}
where in conventions similar to those in \cite{sgiibv1, sgiibv2},
\begin{eqnarray}
\omega = -{\bf{e}}^1\wedge {\bf{e}}^6 - {\bf{e}}^2\wedge {\bf{e}}^7 - {\bf{e}}^3\wedge {\bf{e}}^8-{\bf{e}}^4\wedge {\bf{e}}^9~,
\end{eqnarray}
is an almost Hermitian structure  on ${\cal{S}}$.
Also, noting that
\begin{eqnarray}
\Delta = -2 r^{-2} {Z_+ \over Z_-}~,
\end{eqnarray}
one finds
\begin{eqnarray}
\label{dexp1}
\Delta &=& {1 \over 6} Y_{\ell_1 \ell_2 \ell_3} Y^{\ell_1 \ell_2 \ell_3}
-{1 \over 4} h^2
\nonumber \\
&+& Y_{\ell n_1 n_2} Y^\ell{}_{n_3 n_4}
\bigg[{1 \over 8} \omega\wedge \omega -{1 \over 4} {p \bar{q} \over |p|^2+|q|^2} \chi
-{1 \over 4} {{\bar{p}} q \over |p|^2+|q|^2} {\bar{\chi}} \bigg]^{n_1 n_2 n_3 n_4}~,
\end{eqnarray}
where, in the conventions of \cite{sgiibv1, sgiibv2}
\begin{eqnarray}
\chi = ({\bf{e}}^1+i {\bf{e}}^6)\wedge ({\bf{e}}^2+i{\bf{e}}^7) \wedge ({\bf{e}}^3+i {\bf{e}}^8)\wedge ({\bf{e}}^4+i {\bf{e}}^9)~,
\end{eqnarray}
is the $(4,0)$ form on ${\cal{S}}$.

In particular, on defining
\begin{eqnarray}
\label{dexp2}
{\hat{Y}}_{\ell_1 \ell_2 \ell_3}= (Y_{(0,3)}+Y_{(3,0)})_{\ell_1 \ell_2 \ell_3}
-{i \over 8 (|p|^2+|q|^2)} Y_{m n_1 n_2} \omega^{n_1 n_2}
\bigg( p \bar{q} \chi^m{}_{\ell_1 \ell_2 \ell_3} - {\bar{p}} q {\bar{\chi}}^m{}_{\ell_1 \ell_2 \ell_3} \bigg),
\nonumber \\
\end{eqnarray}
it is straightforward to show, using ({\ref{hexp}}), that ({\ref{dexp1}}) can be rewritten as
\begin{eqnarray}
\label{dexp3}
\Delta = {2 \over 3} {\hat{Y}}_{\ell_1 \ell_2 \ell_3} {\hat{Y}}^{\ell_1 \ell_2 \ell_3}~,
\end{eqnarray}
so $\Delta \geq 0$, as expected.

Next, we consider the remaining components of the KSE. These imply that
\begin{eqnarray}
\label{kse2}
{\tilde{\nabla}}_i \eta_+ -{1 \over 4} h_i \eta_+ -{i \over 12} Y_{\ell_1 \ell_2 \ell_3}\Gamma^{\ell_1 \ell_2 \ell_3} \Gamma_i \eta_+=0~,
\end{eqnarray}
and
\begin{eqnarray}
\label{alg6a}
\bigg( \big[{1 \over 4} {\tilde{\nabla}}_j h_i -{1 \over 8} h_i  h_j +{1 \over 4} Y_{i q_1 q_2} Y_j{}^{q_1 q_2} \big] \Gamma^j
+ \big[{i \over 12}({\tilde{\nabla}}_i Y_{\ell_1 \ell_2 \ell_3}- (dY)_{i \ell_1 \ell_2 \ell_3})
\nonumber \\
+{i \over 24} \big( (h \wedge Y)+ \star_8 (h \wedge Y) \big)_{i \ell_1 \ell_2 \ell_3}
-{1 \over 144} Y_{i m_1 m_2} Y_{m_3 m_4 m_5} \epsilon^{m_1 m_2 m_3 m_4 m_5}{}_{\ell_1 \ell_2 \ell_3}
\nonumber \\
-{1 \over 4} Y_{m [\ell_1 \ell_2} Y_{\ell_3] i}{}^m \big] \Gamma^{\ell_1 \ell_2 \ell_3} \bigg) \eta_+=0~,
\nonumber \\
\end{eqnarray}
where ${\tilde{\nabla}}$ denotes the Levi-Civita connection on ${\cal{S}}$.
Note that on contracting ({\ref{alg6a}}) with $\Gamma^i$, and on making use of ({\ref{alg2}}), one obtains ({\ref{ein1}}).
Furthermore, ({\ref{ein2}}) is obtained from the integrability conditions of the KSE.

Also, on expanding out ({\ref{kse2}}),
one obtains the conditions:
\begin{eqnarray}
\label{kse2a}
\partial_\alpha p + \big({1 \over 2} \Omega_{\alpha, \beta}{}^\beta - iY_{\alpha \beta}{}^\beta -{1 \over 4} h_\alpha
\big)p &=&0~,
\nonumber \\
\partial_\alpha {\bar{p}} + \big(-{1 \over 2} \Omega_{\alpha, \beta}{}^\beta -{1 \over 4} h_\alpha \big){\bar{p}}
-{i \over 3} \epsilon_{\alpha \lambda_1 \lambda_2 \lambda_3} Y^{\lambda_1 \lambda_2 \lambda_3} {\bar{q}} &=&0~,
\nonumber \\
\partial_\alpha q + \big(-{1 \over 2} \Omega_{\alpha, \beta}{}^\beta -{1 \over 4} h_\alpha \big)q +{i \over 3}
\epsilon_{\alpha \lambda_1 \lambda_2 \lambda_3} Y^{\lambda_1 \lambda_2 \lambda_3} p &=&0~,
\nonumber \\
\partial_\alpha {\bar{q}} + \big({1 \over 2} \Omega_{\alpha, \beta}{}^\beta + iY_{\alpha \beta}{}^\beta -{1 \over 4} h_\alpha
\big) {\bar{q}} &=&0~,
\end{eqnarray}
and
\begin{eqnarray}
\label{kse2b}
\Omega_{\alpha, \lambda_1 \lambda_2} \epsilon^{\lambda_1 \lambda_2}{}_{\bar{\mu}_1 \bar{\mu}_2}
 &=& {4 p {\bar{q}} \over |p|^2+|q|^2} \Omega_{\alpha, \bar{\mu}_1 \bar{\mu}_2}~,
\nonumber \\
i Y_{\alpha \bar{\mu}_1 \bar{\mu}_2} -i \delta_{\alpha [\bar{\mu}_1} Y_{\bar{\mu}_2] \beta}{}^\beta
&=& {(|p|^2-|q|^2) \over 2 (|p|^2+|q|^2)} \Omega_{\alpha, \bar{\mu}_1 \bar{\mu}_2} \ .
\end{eqnarray}

So far, we have investigated the general supersymmetric near horizon geometries. From now on, we shall restrict ourselves to some special cases which depend on the choice
of the functions $p$ and $q$ and of the spinor $\eta_+$ in (\ref{pqe}). There are three cases to consider as follows:

\begin{itemize}

\item  $\eta_+$ is an $Spin(7)$ invariant  spinor, $|p|^2=|q|^2$.

\item  $\eta_+$ is a generic $SU(4)$ invariant  spinor,  $p \neq 0$ and $q \neq 0$ and $|p|^2-|q|^2 \neq 0$.

\item  $\eta_+$ is a pure  $SU(4)$ invariant  spinor, $p =0$ or  $q = 0$.

\end{itemize}

We shall investigate in detail the geometry of the spatial horizon section in the first and last cases.

\subsection{$Spin(7)$ invariant Killing spinors}

For solutions with $|p|^2=|q|^2$, one can, by an appropriate $r,u$-independent $Spin(8)$ gauge
transformation, take $q=p$. The conditions on the fields derived from the KSEs can be organized
in $Spin(7)$ irreducible representations but for the analysis that follows it suffices to use their local expressions in $SU(4)\subset Spin(7)$ representations as stated in the previous section. Moreover observe that the 3-form null Killing spinor bi-linear \cite{sgiibv1} which contains the Hermitian 2-form vanishes in this case.

Note first that ({\ref{kse2b}}) implies that the (2,1)  and (1,2) parts of $Y$
vanish, and ({\ref{hexp}}) implies that $h=0$.
Also, from ({\ref{kse2a}}) one finds that $p$ is constant and the $(3,0)$ and $(0,3)$ parts of $Y$ also vanish.
It then follows from ({\ref{dexp1}}) that $\Delta=0$ as well.
Hence, without loss of generality we have
$\epsilon = \eta_+ = 1+e_{1234}$ and $\Delta=0$, $h=0$, $F=0$.
The spacetime is $\mathbb{R}^{1,1} \times {\cal{S}}$, where ${\cal{S}}$ is a compact $Spin(7)$ holonomy manifold.

\subsection{Pure $SU(4)$ invariant Killing spinor}

To analyse these solutions, first note that $\eta_+=p\,1$ is related to $\eta_+=q\, e_{1234}$
by a $r,u$-independent $Spin(8)$ gauge transformation, hence without loss of generality,
it suffices to consider $\eta_+ = p \,1$. Furthermore, an appropriately chosen $u,r$-independent
$U(4)$ gauge transformation can be used  to set $p $ to be a real function. As $|p|^2$ is constant, we can without loss
of generality take
\begin{eqnarray}
\eta_+ =1 \ .
\end{eqnarray}
Then the conditions ({\ref{kse2a}}) are equivalent to
\begin{eqnarray}
\label{aux3a}
Y_{\alpha_1 \alpha_2 \alpha_3}=0~,~~~
\Omega_{\alpha, \beta}{}^\beta -i Y_{\alpha \beta}{}^\beta =0~,~~~
i Y_{\alpha \beta}{}^\beta +{1 \over 2} h_\alpha =0~,
\end{eqnarray}
so, in particular, the $(3,0)$ and $(0,3)$ components of $Y$ vanish.
As $q=0$ as well, it follows from ({\ref{dexp3}}) that
\begin{eqnarray}
\Delta =0 \ .
\end{eqnarray}
Also, ({\ref{kse2b}}) can be rewritten as
\begin{eqnarray}
\label{aux3aa}
\Omega_{\alpha, \lambda_1 \lambda_2} =0~,~~~
i Y_{\alpha \bar{\mu}_1 \bar{\mu}_2} -i \delta_{\alpha [\bar{\mu}_1} Y_{\bar{\mu}_2] \beta}{}^\beta
={1 \over 2}\Omega_{\alpha, \bar{\mu}_1 \bar{\mu}_2} \ .
\end{eqnarray}
Note that these conditions are sufficient to imply that
\begin{eqnarray}
\big( {1 \over 4} h_i \Gamma^i +{i \over 12} Y_{\ell_1 \ell_2 \ell_3} \Gamma^{\ell_1 \ell_2 \ell_3} \big) \eta_+=0~,
\end{eqnarray}
so the Killing spinor is
\begin{eqnarray}
\epsilon = \eta_+ = 1 \ .
\end{eqnarray}
Furthermore, the algebraic condition ({\ref{alg6a}}) can be simplified to obtain
\begin{eqnarray}
\label{aux3c}
\big( (dh)_{ij} \Gamma^j +{i \over 3} (dY-h \wedge Y)_{i \ell_1 \ell_2 \ell_3} \Gamma^{\ell_1 \ell_2 \ell_3}
\big) \eta_+=0~.
\end{eqnarray}
 On contracting ({\ref{aux3c}}) with $\Gamma^i$, and making use of the anti-self-duality of
$dY-h \wedge Y$, one finds
\begin{eqnarray}
(dh)_{ij} \Gamma^{ij} \eta_+=0~,
\end{eqnarray}
i.e.
\begin{eqnarray}
dh_{\alpha \beta}=0, \qquad dh_\alpha{}^\alpha =0~,
\end{eqnarray}
so $dh \in su(4)$. The remaining content of ({\ref{aux3c}}) can be written as
\begin{eqnarray}
dh_{ij} = -(dY-h \wedge Y)_{ijmn} \omega^{mn}~.
\end{eqnarray}

To summarize,  the KSE implies that ${\cal{S}}$ is a Hermitian manifold with  an $SU(4)$ structure
associated with the pair $(\omega, \chi)$ of a Hermitian form $\omega$ and (4,0)-form $\chi$. In addition,
the KSE imposes the geometric condition
\begin{eqnarray}
\theta_\omega=\theta_{\mathrm{Re}\chi}~,
\label{thth}
\end{eqnarray}
where
\begin{eqnarray}
\theta_{\mathrm{Re}\chi}=-{1 \over 4} \star_8 \bigg( \mathrm{Re}\chi \wedge \star_8 d\, \mathrm{Re}\chi \bigg)~,~~~(\theta_\omega)_i = - \nabla^k \omega_{kj} \omega^j{}_i~,
\end{eqnarray}
are the Lee forms of $\mathrm{Re}\chi$ and $\omega$, respectively. This follows on comparing the second equation in (\ref{aux3a}) with the second equation in (\ref{aux3aa}).
Observe that (\ref{thth}) can also be written
as
\begin{eqnarray}
d_{\theta_\omega} \mathrm{Re} \chi=[d\, \mathrm{Re}\chi - \theta_\omega \wedge \mathrm{Re}\chi] =0~.
\end{eqnarray}
We have not included the condition  that $d \theta_\omega \in su(4)$ as this follows from the Hermitian structure on ${\cal S}$.
We shall produce a proof for this in the next section.
Moreover, the components of the metric and fluxes are given as
\begin{eqnarray}
\label{yexp}
\Delta=0, \qquad h= \theta_\omega, \qquad Y = {1 \over 4} (d \omega-\theta_\omega \wedge \omega)~.
\end{eqnarray}
This concludes the analysis of the KSEs.

It remains to investigate the field equations and Bianchi identities. The Bianchi identity $dF=0$  implies that
\begin{eqnarray}
d \star_8 \bigg(d \omega - \theta_\omega \wedge \omega \bigg)=0~.
\label{bianchi}
\end{eqnarray}
The rest of the field equations are also satisfied as a consequence of  (\ref{bianchi}) and the conditions derived from the KSEs.

\subsubsection{Solutions with $\theta_\omega=0$}

Before examining the pure spinor solutions in greater detail, it is instructive to briefly consider the special case for
which $\theta_\omega=0$. The Bianchi identity ({\ref{bianchi}}) implies
that
\begin{eqnarray}
\omega \wedge d \star_8 d \omega =0 \ ,
\label{bal}
\end{eqnarray}
and on integrating this expression over ${\cal{S}}$, one finds that $d \omega =0$, so from
({\ref{yexp}}) it follows that the 5-form flux vanishes, $F=0$, and $\Delta=0, h=0$ so the
spacetime is a product $\mathbb{R}^{1,1} \times {\cal{S}}$, where ${\cal{S}}$ is a compact Calabi-Yau 4-fold.

\newsection{Hidden KT structure of horizon sections}

In this section, we  examine further the properties of the solutions for which the Killing spinor is $\epsilon=1$,
concentrating in particular on the structure of the horizon section ${\cal{S}}$.

\subsection{k-SKT and k-SCYT manifolds}

Before we proceed with the detailed analysis of the geometry of the spatial horizon section, we shall first
explore some geometric structures in the context of 2n-dimensional Hermitian manifolds with Hermitian form $\omega$.  K\"ahler with torsion (KT) manifolds \cite{hkt}
are Hermitian manifolds equipped with the unique compatible
 connection\footnote{In our conventions, we have set $\hat\nabla_i Y^j=\nabla_i Y^j+{1\over2} H^j{}_{ik} Y^k$.}
 $\hat\nabla$ with skew-symmetric torsion $H$, $\hat\nabla\omega=0$.
 Moreover  $H$ is  expressed in terms
 of the complex structure and Hermitian metric as
 \begin{eqnarray}
H=-i_I d\omega=-i(\partial-\bar\partial)\omega~.
\label{htor}
\end{eqnarray}
Clearly $\mathrm{hol}(\hat\nabla)\subseteq U(n)$.
 For strong KT manifolds (SKT),  the torsion is in addition closed, $dH=0$.
 The latter condition can be expressed as
\begin{eqnarray}
\partial\bar\partial \omega=0~.
\end{eqnarray}
 This condition has been extensively investigated in the context of
 supersymmetric 2-dimensional sigma models \cite{hull, sierra, howegp}
and in the context of Hermitian geometry \cite{hkt, poon, ivanovpapv1, ivanovpapv2, anna}.

Another second order equation which arises in the context 2n-dimensional Hermitian manifolds is
\begin{eqnarray}
\partial \bar\partial \omega^{n-1}=0~.
\label{gaud}
\end{eqnarray}
It has been shown by Gauduchon \cite{gauduchon} that within the conformal class of a Hermitian metric, there is a
representative which solves (\ref{gaud}).

To continue,  it is suggestive to define as
 $k$-SKT manifolds the Hermitian manifolds equipped with the compatible connection with skew-symmetric torsion, $H$,
 which in addition satisfies
 \begin{eqnarray}
 d(\omega^{k-1}\wedge H)={2i\over k} \partial\bar\partial\omega^{k}=0~.
 \label{kskt}
 \end{eqnarray}
Clearly for $k=1$ this condition coincides with SKT,  while for a 2n-dimensional Hermitian manifold and for $k=n-1$ it coincides
with the Gauduchon condition (\ref{gaud}).

Next let us compare the above conditions for Hermitian manifolds of different dimension. It is clear that for 4-dimensional
Hermitian manifolds the SKT condition
coincides with the Gauduchon condition, and so all 4-dimensional Hermitian manifolds are SKT.
In 6 dimensions, the 2-SKT condition (\ref{pbpo}) coincides with the Gauduchon condition \cite{gauduchon}. Therefore all 6-dimensional
Hermitian manifolds
are 2-SKT. However, it is known that the SKT condition is restrictive  for 6-dimensional  manifolds
\cite{tian}. It is likely that this
is also the case for the SKT and 2-SKT conditions for 8-dimensional Hermitian manifolds. In this case, the Gauduchon condition coincides
with the 3-SKT structure. Similar observations can be made for Hermitian manifolds in higher dimensions.
The conditions (\ref{kskt}) provide
a set of natural second order equations on Hermitian manifolds which may deserve further investigation.

Next consider CYT manifolds, i.e.~KT manifolds which in addition have $\mathrm{hol}(\hat\nabla)\subseteq SU(n)$.
Clearly the k-strong condition also generalizes in this case yielding a k-SCYT structure.
It is known that there are restrictions on the existence of such manifolds. As an example, closed, conformally balanced, ie
$\theta_\omega=2d\Phi$ and $\Phi$ is a smooth real function, SCYT manifolds
 are Calabi-Yau \cite{ivanovpapv1, ivanovpapv2}.
It is not known under which conditions  similar  theorems hold for k-SKT manifolds, $k\geq 1$.
It turns out that the spatial horizon sections ${\cal S}$ admit a 2-SCYT structure. Moreover we shall provide compact 8-dimensional examples with this structure.
However in all examples, we shall construct  manifolds which are not conformally balanced.

\subsection{Hidden torsion}

Returning to the geometry of the spatial horizon sections, we have shown that   ${\cal{S}}$ is a Hermitian manifold with a  $SU(4)$ structure associated
with the pair $(\omega, \chi)$ of fundamental
forms. In addition, the Killing spinor equations impose the geometric constraint given
 in (\ref{thth}). It turns out that (\ref{thth}) is equivalent to requiring
that ${\cal S}$ is a KT manifold with a compatible  $SU(4)$-structure\footnote{The classes of $SU(3)$-structures
on 6-dimensional manifolds have been investigated in \cite{salamon}.}, i.e.~a CYT manifold.   This has been first  observed for 6-dimensional manifolds  with
an $SU(3)$-structure in \cite{lustb}, and later
it has been
expressed in the form (\ref{thth}) for all 2n-dimensional manifolds with a $SU(n)$-structure  in \cite{hetv1, hetv2}.
This means that there exists a connection with skew-symmetric torsion $H$ such that
\begin{eqnarray}
\hat\nabla \omega=\hat\nabla\chi=0~,
\label{hoc}
\end{eqnarray}
where    $H$ is given in (\ref{htor}).

The 3-form $H$ is not immediately identifiable with either the NS-NS or the R-R 3-form field strengths of IIB supergravity
as we have set both of them to zero. In addition,
$H$ may not be closed and, for a non-product near horizon geometry, ${\cal S}$ should not be balanced, $\theta_\omega\not=0$.

The KSE requires that $d\theta_\omega\in \mathfrak{su}(4)$. To show that the (2,0) part of $d\theta_\omega$ vanishes we can utilize
the existence of $H$ and in particular  (\ref{hoc}). For this first observe
that the Ricci form $\hat\rho$ of $\hat\nabla$ for any KT manifold can be written \cite{ivanovpapv1, ivanovpapv2} as
\begin{eqnarray}
\hat\rho\equiv-{1\over4} \hat R_{ij, k\ell}\, \omega^{k\ell}\, e^i\wedge e^j=-i\partial\bar\partial\log \det g-d(I\theta_\omega)~,
\end{eqnarray}
where $(I\theta_\omega)_i= (\theta_\omega)_j I^j{}_i$. To establish the above identity,
it is convenient to use complex coordinates. Since the holonomy\footnote{It turns out that $d\theta_\omega^{2,0}=0$  for
all Hermitian manifolds, i.e.~$\mathrm{hol}(\hat\nabla)\subseteq U(n)$, but a proof is more involved.}
of $\hat\nabla$ is contained in $SU(4)$,
$\hat\rho=0$. Taking the (2,0) part of the rhs, one finds that $d\theta_\omega^{2,0}=0$. It remains to show that
$d(\theta_\omega)_{ij} \omega^{ij}=0$. This follows from the definition of $\theta_\omega$ and
\begin{eqnarray}
{1\over2}\omega^{ij} (d\theta_\omega)_{ij}=-\omega^{ij} \nabla_i\big(\nabla^k\omega_{k\ell} \omega^\ell{}_j \big)=
\nabla_i\nabla_j \omega^{ij}
+\nabla_k\omega^{ki} \nabla_\ell \omega^{\ell j} \omega_{ij}=\nabla_i\nabla_j \omega^{ij}=0~,
\end{eqnarray}
where one establishes  the last equality by expressing the two derivatives  in terms of the Riemann curvature and
by using that the Ricci tensor is symmetric.

Another advantage of introducing the torsion $H$ is that the Bianchi identity for $F$ (\ref{bianchi}) can now be expressed as
\begin{eqnarray}
d\star_8[d\omega-\theta_\omega\wedge \omega]=d(\omega\wedge H)=0~.
\label{dowh}
\end{eqnarray}
Using (\ref{htor}) observe that the above equation can be rewritten as
\begin{eqnarray}
\partial\bar\partial \omega^2=0~.
\label{pbpo}
\end{eqnarray}
Clearly this is a second order equation on the Hermitian form $\omega$ and it coincides with the 2-strong condition on KT manifolds.

To summarize, both the KSEs and field equations require that spatial horizon section  ${\cal S}$  is a 2-SCYT manifold.
To find examples of  IIB horizons, it is convenient to utilize the hidden torsion of ${\cal S}$ and solve the conditions
required for the 2-SCYT structure. These are two equations, one is the vanishing of the Ricci form $\hat\rho=0$ of the
connection with torsion and the other is the 2-strong condition (\ref{pbpo}).
 There are two  sources of examples of such manifolds.
 One  source is the  Nil-manifolds. However this class
will not produce interesting examples as it has been shown that all  Nil-manifolds with invariant  Hermitian structure and
 $\mathrm {hol} (\hat\nabla)\subseteq SU(4)$ are balanced \cite{anna}.
Since in this case $h=\theta_\omega$ and $\Delta=0$, the near horizon geometry is a product $\mathbb{R}^{1,1}\times {\cal S}$, where ${\cal{S}}$
is a compact Calabi-Yau 4-fold, and the 5-form flux vanishes. In fact as a consequence of the argument given in  section 3.2.1, all
balanced, $\theta_\omega=0$,  2-SKT 8-dimensional manifolds are K\"ahler. The other
source of examples are group fibrations over Hermitian manifolds. We shall demonstrate that this class produces many
examples.

\newsection{KT fibrations}

In this section, we present a number of examples of near-horizon geometries corresponding to
the class of solutions for which the Killing spinor is $\epsilon=1$ by constructing horizon sections satisfying
the conditions described in section 4. As we have shown, the entire near-horizon solution is
completely determined in terms of that of the spatial horizon  section ${\cal{S}}$. Thus we have to find examples of 8-dimensional 2-SCYT manifolds.
For this, we shall consider group fibrations over KT manifolds.

Our primary interest is in 8 dimensions but the construction of fibrations can be made for any 2n-dimensional KT manifold $X^{2n}$.
 To continue suppose that $X^{2n}$ is a  fibration of a group $G$ over a KT 2m-dimensional manifold $B^{2m}$ with metric $ds_{(2m)}^2$,  complex structure $I$ and skew-symmetric
torsion 3-form $H_{(2m)}$. For $G$ a torus such fibrations have been extensively investigated in \cite{sethi, gold, grant} and have been further explored in \cite{yauv1, yauv2}. Here we shall extend the
construction to more general  group fibrations.  For this, we take $X^{2n}=G\times_K P(K, B^{2m})$, ie $X^{2n}$ is $G$ group fibration
associated to a principal fibration $P(K, B^{2m})$, $K\subset G$. In addition, $P(K, B^{2m})$ is equipped with a principal bundle connection $\lambda^A$  and $K$ acts on $G$ from the right.  Considering a metric $h$ on $G$ which is left invariant and assuming that in addition is invariant under the right action of $K$, one can introduce
  a metric and a 3-form   on
$X^{2n}$ as
\begin{eqnarray}
&&ds^2_{(2n)}=h_{ab} \lambda^a \lambda^b+ ds^2_{(2m)}~,~~~~ds^2_{(2m)}=\delta_{ij} e^i e^j~,
\cr
&&H_{(2n)}= {1\over3!} H_{abc} \lambda^a \wedge \lambda^b\wedge \lambda^c+ h_{ab} \lambda^a\wedge {\cal F}^b+ H_{(2m)}~,
\label{kts}
\end{eqnarray}
where the frame $\lambda^a=e^a-\xi^a{}_A \lambda^A$, $e^a$ are the left-invariant 1-forms on $G$,  $\xi$'s are the (left-invariant) vector fields that generate the right $K$ action on $G$, and $H_{abc}$ are the structure constants of $G$. Observe that the $\xi$'s are constant when evaluated on the left invariant frame $e^a$. This construction gauges the right action of $K$ on $G$. Furthermore
\begin{eqnarray}
d\lambda^A-{1\over 2} H^A{}_{BC} \lambda^B\wedge \lambda^C={\cal F}^A~,~~~~{\cal F}^a=\xi_A^a {\cal F}^A~,
\end{eqnarray}
where ${\cal F}^A$ is the curvature of $\lambda^A$ and $H_{ABC}$ are the structure constants of $K$. 
 This relation of $H_{(2n)}$ to the Chern-Simons-like form of $\lambda$ has been motivated by the results
of \cite{hetv1, hetv2}.
 Observe that
\begin{eqnarray}
H_{(2m)}=-i_I d\omega_{(2m)}~,
\end{eqnarray}
where $\omega_{(2m)}$ is the Hermitian form of $B^{2m}$.

To define a KT structure on  $X^{2n}$, we assume that the fibre $G$ admits a left-invariant
almost complex structure $J$ such that $h$ is a Hermitian metric with respect to $J$. In addition, we require that
 the almost Hermitian form is chosen such that it is also invariant under the right action of $K$. This in particular implies that $H^c{}_{Aa} J_{cb}-(b,a)=0$.
Moreover $J$ is chosen such that
the structure constants $H_{abc}$ of the Lie algebra of $G$ are identified with the components of skew-symmetric torsion associated
with the
Hermitian structure $(h, J)$ on $G$ \cite{sevrin}. Using these, one can write an almost Hermitian form on $X^{2n}$ as
\begin{eqnarray}
\omega_{(2n)}={1\over2} J_{ab} \lambda^a\wedge \lambda^b+ \omega_{(2m)}~.
\end{eqnarray}
Next for   $X^{2n}$ to be a complex manifold, one finds the conditions 
\begin{eqnarray}
{\cal F}^a_{ij} I^i{}_k I^j{}_\ell={\cal F}^a_{k\ell}~,
\label{herm1}
\end{eqnarray}
ie the curvature of the fibration is (1,1) with respect to the complex structure of the base space $B^{2m}$,
and
\begin{eqnarray}
H_{abc}-3 H_{ef[a} J^e{}_b J^f{}_{c]}=0~,
\label{herm2}
\end{eqnarray}
ie the structure constants of $G$ are (2,1) and (1,2) with respect to $J$. Thus provided (\ref{herm1}) and (\ref{herm2})
are satisfied, $X^{2n}$  is a KT manifold  with respect to $(ds^2_{(2n)}, \omega_{(2n)})$ with torsion given in (\ref{kts}).

The conditions (\ref{herm1}) and (\ref{herm2}) can be solved as follows. First (\ref{herm2}) is automatically
satisfied because $J$ is chosen such that $H_{abc}$  is the skew-symmetric torsion of the Hermitian structure
$(h,J)$ of $G$. The condition (\ref{herm1}) can be solved by taking the fibration to be holomorphic. Therefore,
$X^{2n}$ is a holomorphic fibration over a Hermitian manifold $B^{2m}$, with fibre $G$ which also admits an invariant
Hermitian structure with skew-symmetric torsion constructed from the structure constants of $G$.

Next for  $X^{2n}$ to have a CYT structure, it is required that the connection with skew-symmetric torsion
 has holonomy contained $SU(n)$, $\mathrm{hol}(\hat\nabla)\subseteq SU(n)$.
Since by construction $\hat\nabla$ preserves both the metric $ds^2_{(2n)}$  and $\omega_{(2n)}$, clearly the holonomy of $\hat\nabla$ is contained in $U(n)$. It remains
to further restrict the holonomy to $SU(n)$.  For this, we set the Ricci form of the connection with skew-symmetric
torsion to zero, $\hat \rho_{(2n)}=0$. This in turn gives the
conditions
\begin{eqnarray}
(\hat \rho_{(2m)})_{k\ell}+{1\over2}h_{ab} {\cal F}^a_{k\ell} {\cal F}^b_{ij} \omega_{(2m)}^{ij}&=&0~,
\cr
2 {\cal F}^a_{ik} {\cal F}^b_{j\ell} \delta^{k\ell} \omega_{(2m)}^{ij}+H^{ab}{}_c {\cal F}^c_{ij} \omega_{(2m)}^{ij}&=&0~,
\cr
\hat\nabla_k({\cal F}^a_{ij} \omega_{(2m)}^{ij})&=&0~,
\end{eqnarray}
where $\hat \rho_{(2m)}$ is the Ricci form of the connection with torsion of $B^{2m}$.
It is clear that
\begin{eqnarray}
{\cal F}^a_{ij} \omega_{(2m)}^{ij}=k^a~,
\label{kkk}
\end{eqnarray}
is constant.
Using this and  that ${\cal F}$ is a (1,1)-form, the above conditions can be simplified somewhat to
\begin{eqnarray}
(\hat \rho_{(2m)})_{k\ell}+{1\over2}h_{ab} k^b {\cal F}^a_{k\ell} &=&0~,
\cr
H^{ab}{}_c k^c&=&0~.
\label{fcon}
\end{eqnarray}
It is clear from the last condition above that if $k\not=0$,  the direction along $k$ in the Lie algebra of $G$ commutes
with all other generators of $G$. Thus up to a discrete identification, $G=U(1)\times G'$.
Finally, one can compute the Lee form to find that
\begin{eqnarray}
(\theta_{\omega_{(2n)}})_i&=&(\theta_{\omega_{(2m)}})_i~,
\cr
(\theta_{\omega_{(2n)}})_a&=&{1\over 2} H_{b_1b_2 c} J^{b_1 b_2} J^c{}_a+{1\over2} k_c J^c{}_a~.
\label{holsu}
\end{eqnarray}
Observe that the first term in the second equation of (\ref{holsu}) is the Lee form associated with
the Hermitian structure $(h,J)$ of $G$. This completes the general analysis on group fibrations and KT structures.

Next take ${\cal S}=X^8$. Since ${\cal S}$ is a CYT manifold   both the fibre group and the base
manifold $B^{2m}$ are restricted. First
the fibre groups  are restricted to be KT manifolds, and with skew-symmetric torsion
obtained from the structure constants of the associated Lie algebra. Furthermore, the fibre groups must admit
a left-invariant  metric and a left-invariant  Hermitian form which are in addition invariant under
 the right action of a subgroup $K$. In the examples explored below $K$ is chosen either
 as the trivial subgroup or a torus.  It turns out  that all even-dimensional compact Lie groups
satisfy
all these conditions. We have tabulated all such groups up to  dimension 8  in table 1. These are relevant for the construction
of horizons.

\begin{table}[ht]
 \begin{center}
\begin{tabular}{|c|c|}
\hline
$\mathrm{dim}\,G$&$G$  \\
\hline
\hline
$2$  & $T^2$ \\
\hline
$4$  & $T^4$, $S^1\times SU(2)$\\
\hline
$6$  & $T^6$, $T^3\times SU(2)$, $SU(2)\times SU(2)$\\
\hline
$8$  & $T^8$, $T^5\times SU(2)$, $T^2\times SU(2)\times SU(2)$, $SU(3)$\\
\hline
\end{tabular}
\end{center}
\label{ttt}
\caption{\small
The first column gives the rank of the fibre which is the dimension of the group. The second column gives the available compact Lie groups up to discrete identifications.}
\end{table}

The only restriction on the fibre group arises whenever the fibre twists over the base space with a connection $\lambda$ such that $k$ in (\ref{kkk})
does not vanish. As we have mentioned in such a case $G$ is a product $U(1)\times G'$ up to a discrete identification. To find new horizon geometries,
it remains to solve for (\ref{fcon}) and (\ref{kkk}), and in addition verify the 2-strong condition $d(\omega_{(8)}\wedge H_{(8)})=0$. We shall do this explicitly in some special cases below. In all the examples below, the requirement that the
metric and Hermitian form to be invariant under the right action of the subgroup $K$ of $G$ that it is gauged is always satisfied.

\subsection{Group Manifold Horizon Sections}

Let us suppose that the horizon section is a group manifold. The $T^8$ case is trivial. Next consider the case $T^5\times SU(2)$ and
take
\begin{eqnarray}
&&ds^2_{(8)}= \sum_{r=1}^5 ( \tau^r)^2 +(\sigma^1)^2+(\sigma^2)^2+(\sigma^3)^2~,
\cr
&&\omega_{(8)}=-\sigma^3 \wedge\tau^1 - \sigma^1\wedge \sigma^2+ {1\over2} \sum_{r,s=2}^5 J_{rs}\tau^r\wedge \tau^s~,
\end{eqnarray}
where
\begin{eqnarray}
d\tau^r=0~, ~~~d\sigma^3=\sigma^1\wedge \sigma^2~,
\end{eqnarray}
and cyclically in $1,2$ and $3$, and $J_{rs}$ a constant complex structure in the denoted 4 directions.
In this case $\hat\nabla$ is a parallelizable connection and so the holonomy is $\{1\}$. Moreover
\begin{eqnarray}
H_{(8)}=\sigma^1\wedge \sigma^2\wedge \sigma^3~,~~~\theta_\omega=\tau^1~,
\end{eqnarray}
and the 2-strong condition can be easily verified.

Next consider $T^2\times SU(2)\times SU(2)$. One can take
\begin{eqnarray}
&&ds^2_{(8)}= \sum_{r=1}^2 (\tau^r)^2 +(\sigma^1)^2+(\sigma^2)^2+(\sigma^3)^2+(\rho^1)^2+(\rho^2)^2+(\rho^3)^2~,
\cr
&&\omega_{(8)}=-\sigma^3 \wedge\rho^3 - \sigma^1\wedge \sigma^2-\rho^1\wedge \rho^2-\tau^1\wedge \tau^2
\end{eqnarray}
where
\begin{eqnarray}
d\tau^r=0~, ~~~d\sigma^3=\sigma^1\wedge \sigma^2~,~~~d\rho^3=\rho^1\wedge \rho^2~,
\end{eqnarray}
and cyclically in $1,2$ and $3$. In such case  $\hat\nabla$ is again parallelizable and
\begin{eqnarray}
H_{(8)}=\sigma^1\wedge \sigma^2\wedge \sigma^3+\rho^1\wedge \rho^2\wedge \rho^3~,~~~\theta_{\omega_{(8)}}=-\sigma^3+\rho^3~.
\end{eqnarray}
A short calculation reveals that the 2-strong condition is also satisfied.

It remains to examine $SU(3)$. For this consider the Hermitian structure associated with the bi-invariant
metric of $SU(3)$ and the complex structure given in \cite{sevrin}. The associated connection with skew-symmetric
torsion is the left-invariant parallelizable connection and so $\mathrm{hol}(\hat\nabla)=\{1\}$. But the condition
(\ref{oskt}) is not satisfied.

\subsection{Fibrations over Riemann surfaces}

Suppose that $B^2$ is a Riemann surface. Equations (\ref{kkk}) and (\ref{fcon}) imply that the curvature of $B^2$ is non-negative. Thus $B^2$ is either $T^2$
or $S^2$. Let us focus on the $S^2$ case. The fibre group is 6-dimensional and from table 1 there are 3 different cases to consider. First suppose that
$G=T^6$. In such case one can write
\begin{eqnarray}
&&ds^2_{(8)}=h_{ab} \lambda^a \lambda^b+ ds^2(S^2)
\cr
&& \omega_{(8)}={1\over2} J_{ab} \lambda^a \wedge \lambda^b +\omega_{(2)}(S^2) \ .
\end{eqnarray}
Moreover (\ref{kkk}) implies that
\begin{eqnarray}
{\cal F}^a_{ij}={1\over2}k^a (\omega_{(2)})_{ij}~,
\label{fko}
\end{eqnarray}
where $k$ is constant.
In turn the first condition implies that
\begin{eqnarray}
R_{ij,k\ell}= {|k|^2\over4} (\omega_{(2)})_{ij} (\omega_{(2)})_{k\ell}~,
\end{eqnarray}
as $H_{(2)}=0$. A straightforward computation reveals that
\begin{eqnarray}
H_{(8)}=h_{ab} \lambda^a\wedge {\cal F}^b~,~~~\theta_{\omega_{(8)}}={1\over2} k^b J_{ba} \lambda^a \ .
\end{eqnarray}
Moreover one can easily verify that $d(\omega_{(8)} \wedge H_{(8)})=0$. Thus any rank 6 toroidal fibration over $S^2$ with curvatures
proportional to the K\"ahler form of $S^2$ solves all the conditions. All such manifolds are 2-SCYT.

Next take $G=T^3\times SU(2)$. Again equations (\ref{kkk}) and (\ref{fcon}) imply that $B$ is either $T^2$ or $S^2$. We shall focus on the latter case.
 The second condition in (\ref{fcon}) and (\ref{kkk})  imply that the fibration curvature along the $SU(2)$ directions vanishes.
Thus there is no twisting of $SU(2)$ over the Riemann surface. As a result, we take only the $T^3$ part of the fibre to twist.
Thus we have
\begin{eqnarray}
&&ds^2_{(8)}=h_{ab} \lambda^a \lambda^b+(\lambda^3)^2+ ds^2(S^3)+ds^2(S^2)~,~~~a,b=1,2~,
\cr
&& \omega_{(8)}={1\over2} J_{ab} \lambda^a \wedge \lambda^b -\sigma^3 \wedge\lambda^3 - \sigma^1\wedge \sigma^2+\omega_{(2)}(S^2)
\end{eqnarray}
where
\begin{eqnarray}
ds^2(S^3)=(\sigma^1)^2+(\sigma^2)^2+(\sigma^3)^2~.
\end{eqnarray}

As in the previous case (\ref{kkk}) implies (\ref{fko}) but now $k$ lies along the 3 toroidal directions.
Moreover
\begin{eqnarray}
H_{(8)}=h_{ab} \lambda^a\wedge {\cal F}^b+ \lambda^3\wedge {\cal F}^3+\sigma^1\wedge \sigma^2\wedge \sigma^3~,~~~
\theta_{\omega_{(8)}}= {1\over2} k^b J_{ba} \lambda^a+\lambda^3 + {1 \over 2} k^3 \sigma^3.
\end{eqnarray}
It remains to verify the 2-strong condition $d(\omega_{(8)}\wedge H_{(8)})=0$. This is satisfied provided that
\begin{eqnarray}
{\cal F}^1={\cal F}^2=0~,
\end{eqnarray}
and so $k^1=k^2=0$. Thus the horizon section is $T^2\times S^3\times S^3$, with one of the 3-spheres possibly squashed.
Observed that in both cases above the data are invariant with respect to the right action of subgroup $K$, which is a torus, that it is gauged.

The last case is for $G=SU(2)\times SU(2)$. There are no solutions in this case as one cannot satisfy all conditions in (\ref{fcon}).

 \subsection{Fibrations over K\"ahler-Einstein manifolds}

\subsubsection{Six-dimensional base space}

First we shall examine horizon sections which are $T^2$-fibrations over 6-dimensional KT manifolds.
At the end we shall consider
 the horizon sections which are fibrations over 4-dimensional KT manifolds.
 To simplify the problem further we shall take $B^6$ to be a K\"ahler-Einstein manifold.
 The Ricci form of such manifolds is proportional to the K\"ahler form. Thus the K\"ahler form, up to an overall scale,
  represents
  the first Chern class of the canonical line bundle. Using the K\"ahler-Einstein condition of $B^6$,
  the metric, torsion and Hermitian form of the horizon section can be written as
 \begin{eqnarray}
&&ds^2_{(8)}=(\lambda^0)^2+  (\lambda^1)^2+ ds^2(B)~,~~~~
\cr
&&H_{(8)}= \lambda^0\wedge {\cal F}^0+\lambda^1\wedge {\cal F}^1~,
\cr
&&\omega_{(8)}=-\lambda^0\wedge \lambda^1 +\omega_{(6)}(B)~,~~~d\omega_{(6)}(B)=0~.
\end{eqnarray}
Moreover, we choose the curvature of $\lambda^0$ as
\begin{eqnarray}
{\cal F}^0={k\over6}\, \omega_{(6)} (B)~,
\end{eqnarray}
setting $k^0=k, k^1=0$.
Observe that this  forces the Ricci form of $B^6$ to be positive {\footnote{Our conventions are
chosen so that the Ricci form $\rho$ of a K\"ahler-Einstein manifold with Hermitian form $\omega$ is positive if $\rho = -c \omega$,
for constant $c>0$.}}.
In what follows, we shall specify ${\cal F}^1$, which  is (1,1) and traceless on $B^6$, and
solve the 2-strong condition $d(\omega\wedge H)=0$ for a variety of base manifolds $B^6$.

First take $B^6=\mathbb{C} P^2\times  S^2$. Write
\begin{eqnarray}
\omega_{(6)}= \omega_{\mathbb{C} P^2}+\omega_{S^2}
\end{eqnarray}
where $\omega_{\mathbb{C} P^2}$ and $\omega_{S^2}$ are the Fubini-Study K\"ahler forms on $\mathbb{C} P^2$ and $S^2$, respectively.
Also set
\begin{eqnarray}
{\cal F}^1= p\,  \omega_{\mathbb{C} P^2}+ q\, \omega_{S^2}~.
\end{eqnarray}
Clearly ${\cal F}^1$ is (1,1). Enforcing that ${\cal F}^1$ is traceless, one finds that
\begin{eqnarray}
2p+q=0~.
\end{eqnarray}
Moreover the 2-strong condition $d(\omega_{(8)}\wedge H_{(8)})=0$ implies that
\begin{eqnarray}
{k^2\over12}+2pq+p^2=0~.
\end{eqnarray}
Thus we find that the system has a solution provided that
\begin{eqnarray}
p=\pm {k\over6}~,~~~q=\mp {k\over3}~.
\end{eqnarray}

To give more  examples, observe that the same calculation can be carried out provided that $\mathbb{C} P^2$ is replaced
by any 4-dimensional K\"ahler-Einstein manifold $X_4$ with positive Ricci form. Such manifolds include
$S^2\times S^2$ and the del Pezzo surfaces which arise from blowing up $\mathbb{C} P^2$
on more than two generic points, for the latter see \cite{tian2}.

\subsubsection{Four-dimensional base space}

One can also consider horizons which are fibrations over a 4-dimensional K\"ahler manifold $B^4$. Start  first with torus fibrations.
 In this case,
\begin{eqnarray}
&& ds^2_{(8)}= (\lambda^0)^2+(\lambda^1)^2+(\lambda^2)^2+(\lambda^3)^2+ds^2_{(4)}~,
\cr
&& H_{(8)}=\lambda^0\wedge {\cal F}^0+\lambda^1\wedge {\cal F}^1+\lambda^2\wedge {\cal F}^2+\lambda^3\wedge {\cal F}^3~,
\cr
&& \omega_{(8)}=-\lambda^0\wedge \lambda^1-\lambda^2\wedge \lambda^3+\omega_{(4)}~,
\end{eqnarray}
and $k^0=k$, $k^1=k^2=k^3=0$.
The condition (\ref{dowh}) implies that
\begin{eqnarray}
{\cal F}^2\wedge {\cal F}^2+{\cal F}^3\wedge {\cal F}^3=0~,~~~{\cal F}^0\wedge {\cal F}^0+{\cal F}^1\wedge {\cal F}^1=0~,
\label{ffff}
\end{eqnarray}
and we remark that $k^2=k^3=0$ implies that ${\cal{F}}^2$, ${\cal{F}}^3$ are traceless (1,1) forms on $B^4$,
so the first condition in ({\ref{ffff}}) implies that
\begin{eqnarray}
{\cal{F}}^2= {\cal{F}}^3=0 \ .
\end{eqnarray}
There is a solution for $B^4=S^2\times S^2$ and
\begin{eqnarray}
{\cal F}^0={p\over2} \omega^1_{S^2}+ {q\over2} \omega^2_{S^2}~,~~~k=p+q~,
\cr
{\cal F}^1={\ell\over2} (\omega^1_{S^2}- \omega^2_{S^2})~,~~~\ell^2=pq~.
\label{solvffff}
\end{eqnarray}
Therefore ${\cal S}$ is a product of $T^2$ with a 6-dimensional manifold.

One can also find solutions with fibre $U(1)\times SU(2)$. In this case, one can show that
\begin{eqnarray}
{\cal F}^0\wedge {\cal F}^0=0~.
\label{oof}
\end{eqnarray}
and the rest of the curvatures ${\cal F}$ along $\mathfrak{su}(2)$
must vanish. The condition (\ref{oof}) is rather restrictive since it implies that the self-intersection of the canonical class must
vanish. This can never be satisfied by a K\"ahler-Einstein 4-manifold. However for Ricci flat K\"ahler manifolds
one can take ${\cal F}^0=0$. In such case, the solutions are products. As a result
one finds that up to discrete identifications the horizon sections are either  $S^1\times S^3\times K_3$
or $S^1\times S^3\times T^4$.

\newsection{Uplifted Near-Horizon Geometries}

Another class of solutions can be constructed as lifts to IIB supergravity  of near-horizon geometries in minimal $N=2$, $D=5$ supergravity
derived in \cite{adsfive}. We shall adopt the notation used in \cite{iibfive} where we distinguish
the near-horizon data for the lower dimensional supergravity from those of the higher dimensional theory by adding a subscript indicating the
dimension of the associated space
 as appropriate. In particular,  the near horizon geometry and 1-form gauge potential flux in five dimensions are
\begin{eqnarray}
&&ds_{(5)}^2 = -r^2 \Delta_{(3)}^2 du^2 + 2 du dr +2r h_{(3)} du + ds^2({{{\cal{S}}}^3})~,
\cr
&&A_{(5)} = {\sqrt{3} \over 2} r \Delta_{(3)} du +a~,
\end{eqnarray}
where $h_{(3)}$,  $a$ and $\Delta_{(3)}$ depend only on the coordinates of the 3-dimensional spatial horizon section ${{\cal{S}}}^3$  and in addition
\begin{eqnarray}
da = -{\sqrt{3} \over 2} \star_3 (h_{(3)} + 2 \ell^{-1} Z^1) \ .
\label{da}
\end{eqnarray}
Moreover, we have equipped  ${{\cal{S}}}^3$ with a basis of 1-forms  $( Z^1, Z^2, Z^3)$ such that $d\mathrm{vol}({{\cal{S}}}^3)=Z^1 \wedge Z^2 \wedge Z^3$, $\ell$ is a nonzero constant and
 $\star_3$ denotes the Hodge dual operation on ${{\cal{S}}}^3$.

The basis elements $Z^i$ satisfy
\begin{eqnarray}
{\tilde{\nabla}}_I Z^i_J &=& - {\Delta_{(3)} \over 2} (\star_3 Z^i)_{IJ} + (\gamma_{(3)})_{IJ} (h_{(3)} . Z^i + 3 \ell^{-1} \delta_{i1})
- Z^i_I (h_{(3)})_J
\nonumber \\
&-& 3 \ell^{-1} Z^i_I Z^1_J +2 \sqrt{3} \ell^{-1} \epsilon_{1ij} a_I Z^j_J~,
\end{eqnarray}
where $\gamma_{(3)}$ denotes the metric on ${{\cal{S}}}^3$, ${\tilde{\nabla}}$ is the Levi-Civita connection on ${{\cal{S}}}^3$,
and $h_{(3)}$ satisfies
\begin{eqnarray}
\star_3 d h_{(3)} - d \Delta_{(3)} - \Delta_{(3)} h_{(3)} = 6\ell^{-1} \Delta_{(3)}  Z^1 \ .
\end{eqnarray}
The 2-form field strength of the 5-dimensional solution is
\begin{eqnarray}
F_{(5)} = {\sqrt{3} \over 2} (- \Delta_{(3)} du \wedge dr - r du \wedge d \Delta_{(3)} - \star_3 h_{(3)}) - \sqrt{3} \ell^{-1} \star_3 Z^1 \ .
\end{eqnarray}
After some manipulation, one finds that the uplifted metric and 5-form flux  can be written as {\footnote{Note that
the null basis element ${\bf{e}}^+$ used in the near-horizon geometries described here is
{\it not} the same as the ${\bf{e}}^+$ used in \cite{iibfive}, although ${\bf{e}}^- = dr + r h$ is the same.
If we denote by ${\bf{e}}'^+$ the basis element in \cite{iibfive}, then ${\bf{e}}'^+ = {\bf{e}}^+
-{1 \over 2 r^2 \Delta_{(3)}^2} {\bf{e}}^- +{\ell \over 2 r \Delta_{(3)}}(d \chi_2 + {4 \over \sqrt{3} \ell} a + {2 \over 3}{\cal{Q}})$.}}

\begin{eqnarray}
ds_{(10)}^2 &=& 2 du dr +2r du\, ( h_{(3)} +  \Delta_{(3)} w )+w^2 + ds^2({{{\cal{S}}}^3}) + ds^2(\mathbb{C} P^2)~,
\cr
F_{(10)}&=& \Theta+ \star \Theta~,
\label{iibbh}
\end{eqnarray}
where $\Theta$ has been given in (\ref{5hflux}) and
the 10-dimensional volume form with respect to which the Hodge duality operation is taken is
\begin{eqnarray}
d\mathrm{vol}_{(10)} = -{1 \over 2} {\bf{e}}^+ \wedge {\bf{e}}^- \wedge Z^1 \wedge Z^2 \wedge Z^3
\wedge w \wedge \omega_{\mathbb{C} P^2}
\wedge \omega_{\mathbb{C} P^2} \ .
\end{eqnarray}
The internal manifold is $\mathbb{C} P^2$ with the  Fubini-Study metric
\begin{eqnarray}
\label{yty1}
ds^2(\mathbb{C} P^2) &=& \ell^2 \bigg( d \alpha^2 + \cos^2 \alpha d \beta^2 + \sin^2 \alpha \cos^2 \alpha (d \chi_1+ (\cos^2
\beta - \sin^2 \beta)d \phi)^2
\nonumber \\
&+&4 \cos^2 \alpha \sin^2 \beta \cos^2 \beta d \phi^2 \bigg)~,
\end{eqnarray}
which is K\"ahler-Einstein, and
\begin{eqnarray}
\label{yty2}
w={\ell \over 2}(d \chi_2 + {4 \over \sqrt{3} \ell} a + {2 \over 3}{\cal{Q}})~.
\end{eqnarray}
 In addition,
\begin{eqnarray}
{\cal{Q}} = 3 \cos^2 \alpha (\sin^2 \beta - \cos^2 \beta) d \phi +{3 \over 2}(\sin^2 \alpha - \cos^2 \alpha) d \chi_1~,
\end{eqnarray}
is the potential for the Ricci form of this metric and the  K\"ahler form $\omega_{\mathbb{C} P^2}$ is given by
\begin{eqnarray}
\omega_{\mathbb{C} P^2} = {1 \over 6} \ell^2 d {\cal{Q}} \ .
\label{dq}
\end{eqnarray}
It follows that the  metric of spatial cross-sections of the 10-dimensional horizon geometry and the 3-form $Y$ which determines $F_{(10)}$ are
\begin{eqnarray}
&&ds^2({{\cal{S}}}^8) = w^2 + ds^2({{{\cal{S}}}^3}) + ds^2({\mathbb{C} P^2})~,
\cr
&&Y = -\ell^{-1} Z^1 \wedge Z^2 \wedge Z^3 -{1 \over 4} ( h_{(3)} + 2 \ell^{-1} Z^1
+ \Delta_{(3)} w ) \wedge \omega_{\mathbb{C} P^2}~,
\end{eqnarray}
with
\begin{eqnarray}
\Delta_{(8)} =0, \qquad h_{(8)} = h_{(3)} + \Delta_{(3)} w  \ .
\end{eqnarray}
Note that although $\Delta_{(8)}=0$, there exist near-horizon solutions with $\Delta_{(3)} \neq 0$ (and in fact with
$d \Delta_{(3)} \neq 0$ as well).

It turns out that the Hermitian form on the spatial horizon section is
\begin{eqnarray}
\omega_{(8)} =   Z^1 \wedge w - Z^2 \wedge Z^3 + \omega_{\mathbb{C} P^2} \ .
\end{eqnarray}
From this, it is straightforward to compute the torsion 3-form associated with the black hole uplift solutions, and one finds
that
\begin{eqnarray}
H_{(8)} ={2\over\ell}\bigg( \omega_{\mathbb{C} P^2} + {\ell \over 2}\star_3\big( h_{(3)} + {4 \over \ell} Z^1 \big) \bigg)
\wedge w
- \Delta_{(3)} Z^1 \wedge Z^2 \wedge Z^3 \ .
\end{eqnarray}
After a short computation using previous conditions like (\ref{da}) and (\ref{dq}), one can verify the 2-strong condition
  $ d(\omega_{(8)} \wedge H_{(8)})=0$. Observe that
the above construction can be easily generalized by  replacing $\mathbb{C} P^2$ with another 4-dimensional K\"ahler Einstein manifold.

Explicit examples of 5-dimensional near horizon geometries have been found by explicitly solving for
 $h_{(3)}, a, \Delta_{(3)}$ and the $Z$'s. All known  examples
have 3-dimensional horizon sections ${{\cal{S}}}^3$ which admit two commuting rotational
isometries, which are also symmetries of the full solution. There are three cases of particular interest to consider.

\subsection{Cohomogeneity-2 BPS Black Holes in $D=5$}

The near-horizon geometry of the cohomogenity-2 BPS black holes of Chong et al. \cite{chong} has near-horizon data~\cite{reallk}
\begin{eqnarray}
\label{NHmetricA}
 ds^2_{{{\cal{S}}}^3} &=& \frac{\ell^2 \Gamma d\Gamma^2}{4 P(\Gamma)} + \left( C^2 \Gamma -
 \frac{\Delta_0^2}{\Gamma^2} \right) \left( dx^1 + \frac{\Delta_0  (\alpha_0 - \Gamma)}{C^2 \Gamma^3 - \Delta_0^2} dx^2 \right)^2
 + \frac{4 \Gamma P(\Gamma)}{\ell^2 (C^2 \Gamma^3 - \Delta_0^2)} (dx^2)^2 \ , \nonumber \\
\end{eqnarray}
where
\begin{equation}
P(\Gamma) = \Gamma^3 - \frac{C^2 \ell^2}{4} \left( \Gamma- \alpha_0 \right)^2 - \frac{\Delta_0^2}{C^2}
\end{equation}
with $C$, $\Delta_0$ and $\alpha_0$ constant with $\Delta_0>0$.
Furthermore,
\begin{eqnarray}
 \Delta_{(3)} &=& \frac{\Delta_0}{\Gamma^2}
 \end{eqnarray}
 and
\begin{eqnarray}
 h_{(3)} = \Gamma^{-1} \bigg( \left( C^2 \Gamma - \frac{\Delta_0^2}{\Gamma^2} \right)
  \left( dx^1 + \frac{\Delta_0  (\alpha_0 - \Gamma)}{C^2 \Gamma^3 - \Delta_0^2} dx^2 \right)  - d\Gamma \bigg)
 \end{eqnarray}
and
\begin{equation}
Z^{1} = \frac{\ell(\alpha_0 - \Gamma)C^2}{2\Gamma} dx^{1} + \frac{2\Delta_{0}}{\ell C^2\Gamma} dx^{2} +\frac{\ell}{2 \Gamma} d\Gamma
\end{equation}
and
\begin{eqnarray}
da = -{\sqrt{3} \over 2} \Gamma^{-2} \big(-\Delta_0 dx^1 + \alpha_0 dx^2) \wedge d \Gamma \ .
\end{eqnarray}
From this information, the whole geometric structure associated with the 8-dimensional horizon section ${\cal{S}}$ can be
reconstructed. Note that the Ricci scalar of the metric ({\ref{NHmetricA}}) is not constant, and $h$ does not correspond
to an isometry of ${\cal{S}}$.

\subsection{Cohomogeneity-1 BPS Black Holes in $D=5$}

These were the first examples of supersymmetric, asymptotically $AdS_5$ black holes, with regular horizons.
The near horizon data is as follows; $\Delta_{(3)}$ is a positive constant, and
\begin{eqnarray}
h_{(3)} = - {3 \over \ell} Z^1
\end{eqnarray}
where one can choose that basis $Z^i$ for ${\cal{S}}_3$ satisfying
\begin{eqnarray}
dZ^1 &=& - \Delta_{(3)} Z^2 \wedge Z^3
\nonumber \\
dZ^2 &=& \Delta_{(3)} (1-3 \ell^{-2} \Delta_{(3)}^{-2}) Z^1 \wedge Z^3
\nonumber \\
dZ^3 &=& -\Delta_{(3)} (1-3 \ell^{-2} \Delta_{(3)}^{-2}) Z^1 \wedge Z^2
\end{eqnarray}
with
\begin{eqnarray}
a = -{\sqrt{3} \over 2} \ell^{-1} \Delta_{(3)}^{-1} Z^1
\end{eqnarray}
and it is clear that in this case, ${{\cal{S}}}^3$ is a squashed 3-sphere, and $h_{(8)}$ is a Killing vector on ${\cal{S}}^8$.

\subsection{$AdS_5 \times S^5$}

It is straightforward to write $AdS_5 \times S^5$ as an uplifted solution. The near-horizon data is as follows:
$\Delta_{(3)}=0$, $a=0$, $h_{(3)}= -{2 \over \ell} Z^1$, where the basis $Z^i$ satisfies
\begin{eqnarray}
d Z^1 &=&0
\nonumber \\
dZ^2 &=& \ell^{-1} Z^1 \wedge Z^2
\nonumber \\
dZ^3 &=& \ell^{-1} Z^1 \wedge Z^3 \ .
\end{eqnarray}
Hence, one can introduce local co-ordinates $x, y, z$ such that
\begin{eqnarray}
Z^1= dz, \qquad Z^2= e^{z \over \ell} dx, \qquad Z^3= e^{z \over \ell} dy
\end{eqnarray}
and so the spacetime metric is
\begin{eqnarray}
ds_{(10)}^2 &=& ds^2 (AdS_5)+ ds^2 (S^5)
\end{eqnarray}
where
\begin{eqnarray}
ds^2 (AdS_5) & =&2 du dr -{4r \over \ell} du dz  +dz^2 + e^{2z \over \ell} (dx^2+dy^2),
\nonumber \\ ds^2(S^5)&=& w^2  + ds^2(\mathbb{C} P^2) \ ,
\end{eqnarray}
and $ds^2(\mathbb{C} P^2)$ and $w$ are given by ({\ref{yty1}}) and ({\ref{yty2}}).
The 8-dimensional horizon section is ${{\cal{S}}}^8 = H_3 \times S^5$, where $H_3$ is hyperbolic 3-space.

\newsection{Conclusions}

We have solved the KSEs of IIB near horizon geometries with only 5-form flux preserving at least 2 supersymmetries.
 We demonstrated that there are three cases to consider depending on the choice of Killing spinor which lead to different
 geometries on the spatial horizon sections. We have examined in detail two of these three cases. If the Killing spinor is constructed from a  $Spin(7)$ invariant spinor on the spatial
 horizon section ${\cal S}$, then
 the near horizon geometry is a product $\mathbb{R}^{1,1}\times {\cal S}$, where ${\cal S}$ is an 8-dimensional holonomy $Spin(7)$ manifold.
 For the other  case we investigated, the Killing spinor is constructed from a  $SU(4)$-invariant pure spinor of ${\cal S}$. In this case ${\cal S}$
  is a Hermitian manifold with a $SU(4)$ structure. The most striking property of ${\cal S}$ is that it admits a {\it hidden} K\"ahler with torsion structure
 compatible with the $SU(4)$ structure, i.e.~a Calabi-Yau with torsion structure. The presence of this torsion $H$ is not apparent as both the R-R and NS-NS
3-form field strengths have been set to zero. Moreover, the rotation of the horizon is given by the Lee form of the Hermitian form $\omega$.
 All the remaining equations, including field equations, are also satisfied
provided that $d(\omega\wedge H)=\partial \bar\partial \omega^2=0$. It is therefore clear that the torsion $H$ completely characterizes
the near horizon geometry.

We have utilized the existence of K\"ahler with torsion structure on the spatial horizon sections
to  provide many examples of near horizon geometries mostly constructed from group fibrations over K\"ahler with torsion
manifolds of lower dimension. We also demonstrated that lifted lower-dimensional near horizon geometries to IIB
satisfy all the conditions we have found. Thus there is  a large class of examples.

The condition  $d(\omega\wedge H)=0$ on K\"ahler with torsion manifolds admits various generalizations which  we have explained, like for example  $d(\omega^{k-1}\wedge H)=0$.
We have also compared $d(\omega\wedge H)=0$
with  the strong condition
 $d H=\partial \bar\partial \omega=0$ on K\"ahler manifolds with torsion, which arises in the context of heterotic horizons. The expression for the above condition
 in terms of the torsion allows for a generalization to other manifolds with structure group different from $SU(n)$ which however is compatible with
 a connection with skew-symmetric torsion. We gave a list of several possibilities. It would be of interest to construct examples of manifolds
 satisfying such conditions.

 All the examples of horizons we have constructed so far admit more symmetries than those one a priori  expects to be present in the problem.
 A general solution to the problem will require the solution of two second order differential equations $\hat\rho=0$ and $\partial \bar\partial \omega^2=0$
 on an 8-dimensional complex manifold. The first  involving the Ricci form, $\hat\rho$, of the connection with skew torsion will enforce the condition
 that the associated connection has  (reduced) holonomy  contained in $SU(4)$,
 and the second will enforce the remaining equations of IIB supergravity including field equations. These equations can be contrasted with
 the two equations that arise in the context of heterotic horizons $\hat\rho=0$ and $\partial \bar\partial \omega=0$ as well as the two equations
 that arise in the context of Calabi-Yau manifolds $\rho=0$ and $d \omega=0$, where now $\rho$ is the Ricci form of the Levi-Civita connection. Therefore
 all these manifolds can be viewed as a generalization of Calabi-Yau manifolds.

There is one remaining class of IIB horizons which we have not investigated in this paper. This is associated with a generic $SU(4)$ invariant  spinor on ${\cal S}$.
If solutions exist in this case, the spatial horizon sections are almost complex manifolds but the almost complex structure is not integrable. Moreover, although
the spatial horizon sections have an $SU(4)$ structure, this structure is not compatible with a connection with skew-symmetric torsion.
Therefore, the geometry of the horizons in this case is different from that we have encountered so far in the pure spinor case. We shall examine this case separately
in another publication.

\vskip 0.5cm
\noindent{\bf Acknowledgements} \vskip 0.1cm
\noindent We thank Anna Fino for correspondence and valuable discussions
 on nil-manifolds. GP thanks the Gravitational Physics  Max-Planck Institute at Potsdam for hospitality
where part of this work was done. UG is supported by the Knut and Alice Wallenberg Foundation. JG is supported by the EPSRC grant, EP/F069774/1.
GP is partially supported by the EPSRC grant EP/F069774/1 and the STFC rolling grant ST/G000/395/1.
\vskip 0.5cm

 \setcounter{section}{0}

\appendix{Conventions}

We have used extensively in our calculations that the non-vanishing components of the spin connection associated with the basis ({\ref{basis1}}) are
\begin{eqnarray}
&&\Omega_{-,+i} = -{1 \over 2} h_i~,~~~
\Omega_{+,+-} = -r \Delta, \quad \Omega_{+,+i} =r^2( {1 \over 2}  \Delta h_i -{1 \over 2} \partial_i \Delta),
\cr
&&\Omega_{+,-i} = -{1 \over 2} h_i, \quad \Omega_{+,ij} = -{1 \over 2} r dh_{ij}~,~~~
\Omega_{i,+-} = {1 \over 2} h_i, \quad \Omega_{i,+j} = -{1 \over 2} r dh_{ij},
\cr
&&\Omega_{i,jk}= \tilde\Omega_{i,jk}
\end{eqnarray}
where $\tilde\Omega$ denotes the spin-connection of the 8-manifold ${{\cal{S}}}$ with basis ${\bf{e}}^i$.

In the analysis of the KSE, we have split the spinors $\xi$ into positive and negative  parts as
\begin{eqnarray}
\xi=\xi_+ + \xi_-, \qquad \Gamma_{\pm} \xi_{\pm}=0~.
\end{eqnarray}
Note that if $\xi$ is an even spinor, then
\begin{eqnarray}
\Gamma_{\ell_1 \ell_2 \ell_3 \ell_4} \xi_\pm &=& \pm {1 \over 4!} \epsilon_{\ell_1 \ell_2 \ell_3 \ell_4}{}^{q_1 q_2 q_3 q_4}
\Gamma_{q_1 q_2 q_3 q_4} \xi_\pm~,
\nonumber \\
\Gamma_{\ell_1 \ell_2 \ell_3 \ell_4 \ell_5} \xi_\pm &=& \pm{1 \over 3!} \epsilon_{\ell_1 \ell_2 \ell_3
\ell_4 \ell_5}{}^{q_1 q_2 q_3} \Gamma_{q_1 q_2 q_3} \xi_\pm~,
\nonumber \\
\Gamma_{\ell_1 \ell_2 \ell_3 \ell_4 \ell_5 \ell_6} \xi_\pm
&=& \mp {1 \over 2} \epsilon_{\ell_1 \ell_2 \ell_3 \ell_4 \ell_5 \ell_6}{}^{q_1 q_2} \Gamma_{q_1 q_2} \xi_\pm~,
\nonumber \\
\Gamma_{\ell_1 \ell_2 \ell_3 \ell_4 \ell_5 \ell_6 \ell_7} \xi_\pm
&=& \mp  \epsilon_{\ell_1 \ell_2 \ell_3 \ell_4 \ell_5 \ell_6 \ell_7}{}^{q} \Gamma_{q} \xi_\pm~,
\nonumber \\
\Gamma_{\ell_1 \ell_2 \ell_3 \ell_4 \ell_5 \ell_6 \ell_7 \ell_8} \xi_\pm
&=& \pm \epsilon_{\ell_1 \ell_2 \ell_3 \ell_4 \ell_5 \ell_6 \ell_7 \ell_8} \xi_\pm~,
\end{eqnarray}
whereas if $\xi$ is an odd spinor then
\begin{eqnarray}
\Gamma_{\ell_1 \ell_2 \ell_3 \ell_4} \xi_\pm &=& \mp {1 \over 4!} \epsilon_{\ell_1 \ell_2 \ell_3 \ell_4}{}^{q_1 q_2 q_3 q_4}
\Gamma_{q_1 q_2 q_3 q_4} \xi_\pm~,
\nonumber \\
\Gamma_{\ell_1 \ell_2 \ell_3 \ell_4 \ell_5} \xi_\pm &=& \mp{1 \over 3!} \epsilon_{\ell_1 \ell_2 \ell_3
\ell_4 \ell_5}{}^{q_1 q_2 q_3} \Gamma_{q_1 q_2 q_3} \xi_\pm~,
\nonumber \\
\Gamma_{\ell_1 \ell_2 \ell_3 \ell_4 \ell_5 \ell_6} \xi_\pm
&=& \pm {1 \over 2} \epsilon_{\ell_1 \ell_2 \ell_3 \ell_4 \ell_5 \ell_6}{}^{q_1 q_2} \Gamma_{q_1 q_2} \xi_\pm~,
\nonumber \\
\Gamma_{\ell_1 \ell_2 \ell_3 \ell_4 \ell_5 \ell_6 \ell_7} \xi_\pm
&=& \pm  \epsilon_{\ell_1 \ell_2 \ell_3 \ell_4 \ell_5 \ell_6 \ell_7}{}^{q} \Gamma_{q} \xi_\pm~,
\nonumber \\
\Gamma_{\ell_1 \ell_2 \ell_3 \ell_4 \ell_5 \ell_6 \ell_7 \ell_8} \xi_\pm
&=& \mp \epsilon_{\ell_1 \ell_2 \ell_3 \ell_4 \ell_5 \ell_6 \ell_7 \ell_8} \xi_\pm~.
\end{eqnarray}

\appendix{New geometries with torsion}

As we have seen the second order equation  (\ref{kskt}) on KT manifolds can be expressed in terms of the skew-symmetric
torsion $H$. Because of this it can be extended to other manifolds with a $G$-structure compatible with a
connection with skew-symmetric torsion. We have already investigated the cases with $U(n)$ and $SU(n)$ structures. Here, we
shall explore similar conditions on manifolds with almost KT,  $Sp(n)$, $Sp(n)\cdot Sp(1)$, $G_2$ and $Spin(7)$ structure.

\subsection{k-SAKT manifolds}

Almost KT manifolds (AKT) are almost hermitian manifolds compatible with a connection with skew-symmetric torsion $H$. This have arisen
in the context of supersymmetric 2-dimensional sigma models in \cite{pvnv1, pvnv2}.  In this  case, the expression
for $H$ in terms of the almost Hermitian and almost complex structure  has been given in \cite{stefang2v1, stefang2v2}.
As in the
KT case, we can define as k-SAKT manifolds those spaces for which the AKT structure satisfies the second order equation
\begin{eqnarray}
d(\omega^{k-1}\wedge H)=0~,
\end{eqnarray}
where $\omega$ is the almost Hermitian form.
Unlike the k-SKT condition, the above restriction cannot be easily expressed in terms of a $\partial\bar\partial$ operator. Nevertheless,
it is identical to the k-SKT condition when it is expressed in terms of  $H$.

As has been mentioned in the introduction, one can also define the $(k;\ell)$-SAKT condition as
\begin{eqnarray}
d(\omega^{k-1}\wedge H)\wedge \omega^\ell=0~.
\end{eqnarray}
Clearly this generalizes the k-SAKT structure for $\ell\geq 1$.

\subsection{$(k_1, k_2, k_3)$-SHKT and k-SQKT manifolds}

It is clear that the condition (\ref{kskt}) can easily be extended in the context of HKT manifolds \cite{hkt}, that
is hyper-complex manifolds
equipped with a compatible connection with skew-symmetric torsion.
The expression of the condition (\ref{kskt}) in terms of $H$ naturally leads to an extension of  the strong HKT condition (SHKT)
 to a $(k_1, k_2, k_3)$-SHKT structure as
\begin{eqnarray}
d(\omega^{k_1-1}_I\wedge \omega^{k_2-1}_J\wedge \omega^{k_3-1}_K\wedge H)=0~,
\end{eqnarray}
where $I,J$ and $K$ is a hyper-complex structure and $\omega_I$, $\omega_J$ and $\omega_K$, are the
associated Hermitian forms respectively.
When two of the three $k_1,k_2$ and $k_3$ integers vanish, the above condition coincides with that in (\ref{kskt}). Similarly,
one can define the $(k_1, k_2, k_3; \ell_1,\ell_2,\ell_3)$-SHKT structure as
\begin{eqnarray}
d(\omega^{k_1-1}_I\wedge \omega^{k_2-1}_J\wedge \omega^{k_3-1}_K\wedge H)\wedge \omega^{\ell_1}_I\wedge \omega^{\ell_2}_J\wedge \omega^{\ell_3}_K=0~.
\end{eqnarray}

A similar condition can also be written for QKT manifolds, i.e.~manifolds with a $Sp(n)\cdot Sp(1)$ structure compatible with a connection
with skew-symmetric torsion, \cite{qkt}. In particular, one can define as k-SQKT manifolds the
QKT manifolds which in addition satisfy
 \begin{eqnarray}
d(\psi^{k-1}\wedge H)=0~,
\end{eqnarray}
where
\begin{eqnarray}
\psi=\omega_I\wedge \omega_I+\omega_J\wedge \omega_J+\omega_K\wedge \omega_K~.
\end{eqnarray}
A $(k;\ell)$-SQKT condition can also be defined as
 \begin{eqnarray}
d(\psi^{k-1}\wedge H)\wedge \psi^\ell=0~.
\end{eqnarray}

\subsection{$G_2$ and $Spin(7)$}

 The above conditions can also be extended to manifolds with $Spin(7)$ and $G_2$ structures. It is known that all  8-dimensional manifolds with a $Spin(7)$ structure
 admit a compatible connection with skew-symmetric torsion \cite{stefanspin7}. The torsion $H$ of this connection may not be closed. So one natural second order equation
 on the $Spin(7)$ structure is to impose the closure of $H$, $dH=0$, which is the analogue of the strong condition for SKT manifolds. Alternatively, one can
 impose
 \begin{eqnarray}
 d(\phi\wedge H)=0~,
 \label{spin7}
 \end{eqnarray}
 where $\phi$ is the fundamental self-dual 4-form of the $Spin(7)$ structure.

 A 7-dimensional manifold with a $G_2$ structure  admits a compatible connection with skew-symmetric torsion provided a certain geometric condition is satisfied \cite{stefang2v1, stefang2v2}.
 Again, one can either impose as a second order equation the strong condition, $dH=0$, or alternatively
 \begin{eqnarray}
 d(\varphi \wedge H)=0~,
 \label{g2}
 \end{eqnarray}
 where $\varphi$ is the fundamental 3-form of the $G_2$ structure. Observe that in both the $Spin(7)$ and $G_2$ cases,  the conditions (\ref{spin7}) and (\ref{g2})
 impose a single restriction on the corresponding structures. Both these conditions can be rewritten as ${}^*d{}^*\theta_\phi=0$ and  ${}^*d{}^*\theta_\varphi=0$, where $\theta_\phi$
 and $\theta_\varphi$ are the Lee forms of $\phi$ and $\varphi$,  respectively. So these are Gauduchon type of conditions.

\appendix{Uplifted Horizons}

The  self-dual 5-form of the lifted 5-dimensional black hole solutions is
\begin{eqnarray}
F_{(10)}= \Theta+ \star_{10} \Theta
\end{eqnarray}
where
\begin{eqnarray}
\Theta &=& -{1 \over \ell} {\bf{e}}^+ \wedge {\bf{e}}^- \wedge Z^1\wedge Z^2 \wedge Z^3
+ r \Delta_{(3)} {\bf{e}}^+ \wedge Z^1 \wedge Z^2 \wedge Z^3 \wedge w
\nonumber \\
&+& \omega_{\mathbb{C} P^2}\wedge \bigg({1 \over 4} \Delta_{(3)} Z^1 \wedge Z^2 \wedge Z^3
-{1 \over 4} {\bf{e}}^+ \wedge {\bf{e}}^- \wedge (h_{(3)}+{2 \over \ell} Z^1)
\nonumber \\
&+&{1 \over 4} r {\bf{e}}^+ \wedge \star_3 (-d \Delta_{(3)}+\Delta_{(3)} h_{(3)})
+{1 \over 4}  r {\bf{e}}^+ \wedge w
\wedge (h_{(3)}+{2 \over \ell} Z^1) \bigg)
\nonumber \\
\star_{10} \Theta &=&{\ell\over2} w \wedge \bigg(-{1 \over 4}
\omega_{\mathbb{C} P^2} \wedge \omega_{\mathbb{C} P^2}
-{1 \over 8} \ell \Delta_{(3)} {\bf{e}}^+ \wedge {\bf{e}}^- \wedge \omega_{\mathbb{C} P^2}
\nonumber \\
&-&{1 \over 8} \ell \star_3 (h_{(3)}+{2 \over \ell}Z^1) \wedge \omega_{\mathbb{C} P^2}
+{1 \over 8} \ell r {\bf{e}}^+ \wedge(-d \Delta_{(3)} + \Delta_{(3)} h_{(3)}) \wedge \omega_{\mathbb{C} P^2} \bigg)
\nonumber \\
&+&{1 \over 4} r \Delta_{(3)} {\bf{e}}^+ \wedge \omega_{\mathbb{C} P^2} \wedge \omega_{\mathbb{C} P^2}
-{1 \over 4} r {\bf{e}}^+ \wedge \star_3 (h_{(3)} +{2 \over \ell} Z^1) \wedge \omega_{\mathbb{C} P^2}~.
\label{5hflux}
\end{eqnarray}
This together with the metric in (\ref{iibbh}) describes the full 10-dimensional solution.

\end{document}